\documentclass[11pt]{article} 

\usepackage[PostScript=dvips]{diagrams}

\newcommand{\lub}{\mbox{$\bigsqcup$}}

\newcommand{\category}[1]{\mbox{$\Bbb{#1}$}}

\newtheorem{thm}{Theorem}
\newtheorem{df}{Definition}
\newtheorem{proposition}{Proposition}
\newtheorem{theorem}{Theorem}
\newtheorem{lemma}{Lemma}
\newtheorem{cor}{Corollary}
\newcommand{\epr}{\hfill\raisebox{-.3ex}{\hspace{.2in}\rule{2mm}{3mm}}}
\newcommand{\pr}{{\em Proof: }}
\newcommand{\impl}{\mbox{$\Rightarrow$}}
\newcommand{\ifof}{\mbox{$\Longleftrightarrow$}}

\usepackage{amssymb,stmaryrd}

\newcommand{\llpar}{\bindnasrepma}
\let\lltensor\otimes

\def\llto{\mathbin{-\mkern-3mu\circ}}

\begin{document} 
\bibliographystyle{alpha}
\title{Games and Full Completeness for Multiplicative Linear Logic}
\author{Samson Abramsky and Radha Jagadeesan \\
Department of Computing\\
Imperial College of Science, Technology and Medicine\\
\\
Technical Report DoC 92/24}

\maketitle

\begin{abstract}
We present a game semantics for Linear Logic, in which formulas denote
games and proofs denote winning strategies.  We show that our semantics
yields a categorical model of Linear Logic and prove  {\em full
completeness} for Multiplicative Linear Logic with the MIX rule: every
winning strategy is the denotation of a unique cut-free proof net.  A key
role is played by the notion of {\em history-free} strategy; strong connections
are made between history-free strategies and the Geometry of Interaction.
Our semantics incorporates a natural notion of polarity, leading to a
refined treatment of the additives.  We make comparisons with related work
by Joyal, Blass {\it et al}. 
\end{abstract}

\newpage
\tableofcontents

\newcommand{\ofcourse}{\mbox{$!$}}
\newcommand{\whynot}{\mbox{$?$}}
\newcommand{\with}{\mbox{$\&$}}
\newcommand{\aor}{\mbox{$\oplus$}}
\newcommand{\tensor}{\mbox{$\lltensor$}}
\newcommand{\parc}{\mbox{$\llpar$}}
\newcommand{\cut}[1]{\mbox{$\; \cdot_{#1}\;$}}
\newcommand{\ask}{\mbox{$\rightarrow$}}
\newcommand{\derelict}{\mbox{$D$}}
\newcommand{\weaken}{\mbox{$W$}}
\newcommand{\contract}{\mbox{$C$}}
\newcommand{\emptyvar}{\mbox{\_}}
\newcommand{\tunit}{\mbox{\bf 1}}

\newcommand{\Property}{\mbox{$\cal{P}\;$}}

\newcommand{\mytwotrans}[2]{\mbox{$ \frac{\mbox{$#1$}}{\mbox{$#2$}}$}}
\newcommand{\twoway}{\mbox{$ \rightleftharpoons$}}
\newcommand{\twowayt}{\mbox{$ \stackrel{*}{\twoway} $}}

\newcommand{\sq}[1]{\mbox{$|#1|$}}
\newcommand{\NN}{\mbox{$\cal N$}}
\newcommand{\up}{\mbox{$\uparrow$}}
\newcommand{\DD}{\mbox{$\cal D$}}
\newcommand{\alt}{\mbox{$\;{\tt\char`\|}\;$}}
\newcommand{\fix}[2]{\mbox{$#1\ \underline{{\tt fix}}\  #2$}}

\newcommand{\linimpl}{\mbox{$\llto$}}

\newcommand{\entails}{\mbox{$\;\vdash$}}
\newcommand{\twotrans}[2]{\mbox{$\frac{\mbox{$#1$}}{\mbox{$#2$}}$}}
\newcommand{\eone}[1]{\mbox{$[\![ #1 ]\!]$}}


\newcommand{\Ex}{\mbox{$EX$}}
\newcommand{\Fb}{\mbox{$FB$}}
\newcommand{\conv}{\mbox{$\Downarrow$}}
\newcommand{\tinv}[1]{\mbox{$#1^{-1}$}}
\newcommand{\tauinv}{\mbox{$\tau^{-1}$}}

\newcommand{\Predom}{\mbox{\tt {\bf Predom}}}
\newcommand{\Dom}{\mbox{\tt {\bf Dom}}}
\newcommand{\Games}{\mbox{$\cal{G}$}}
\newcommand{\gimorphism}[2]{\mbox{$#1 \leadsto #2 $}}
\newcommand{\lra}{\mbox{LRA}}

\newcommand{\que}[1]{\mbox{${#1}^{\raise.5ex\hbox{$\scriptscriptstyle \wedge$}}$}}
\newcommand{\ans}[1]{\mbox{${#1}_{\hbox{$\scriptscriptstyle \vee$}}$}}
\newcommand{\Gi}{\mbox{$\cal{G}I$}}

\newcommand{\posg}[1]{\mbox{${#1}^{+}$}}
\newcommand{\negg}[1]{\mbox{${#1}^{-}$}}

\section{Introduction}

We present a Game Semantics for Linear Logic \cite{Gir87}, in 
which 
formulas denote games, and proofs denote winning strategies.
We also prove a novel kind of Completeness Theorem  for this  semantics, which
says that every strategy in the model is the denotation of some proof.

Our motivation is threefold:
\begin{itemize}
\item We believe that the Game Semantics captures the dynamical intuitions
behind Linear Logic better than any other extant semantics.

\item We see Game Semantics as potentially providing a very powerful
unifying framework for the semantics of computation, allowing typed
functional languages, concurrent processes and complexity to
be handled in an integrated fashion.

\item Game Semantics mediates between traditional operational and
denotational semantics, combining the good structural properties
of one with the ability to model computational fine structure of the
other. This is similar to the motivation for the Geometry of Interaction
programme \cite{TGI,GI1,Abr92}; indeed, we shall exhibit strong 
connections between  our semantics
and the Geometry of Interaction.
\end{itemize}

\subsection{Overview of Results}

Blass has recently described a Game semantics for Linear
Logic~\cite{Bla92}. 
This has good claims to be the most intuitively
appealing semantics for Linear Logic presented so far.  However, there is a
considerable gap between Blass' semantics and Linear Logic:
\begin{enumerate}
\item The semantics validates Weakening, so he is actually modelling Affine
logic.  
\item Blass characterises validity in his interpretation for the
multiplicative fragment: a formula is game semantically valid if and only
if it is an instance of a binary classical propositional tautology (where
tensor, par, linear negation are read as classical  conjunction,
disjunction and negation).   
Thus there is a big gap even between provability in Affine logic and validity
in his semantics.
\end{enumerate}
This leaves open the challenge of refining Blass' interpretation to get a
closer fit with Linear Logic, while retaining its intuitive appeal.  

On the other hand, there is the challenge of obtaining a {\em full
completeness
theorem}.  The usual completeness theorems are stated with respect to
provability;  a full completeness theorem is with respect to proofs.
This is best formulated in terms of a categorical model of the
logic, in which formulas denote objects, and proofs denote
morphisms.  One is looking for a model \category{C} such that:
\begin{description}
\item[Completeness: ]
$\category{C}(A,B)$ is non-empty only if $A \entails B$ is provable in the
logic.
\item [Full Completeness: ] 
Any $f: A \rightarrow B$ is the denotation of a proof of $A
\entails B$.  (This amounts to asking that the unique functor from the relevant
free category to \category{C} be full, whence our terminology).  One may
even ask for there to be a {\em unique} cut-free such proof, {\em i.e.}
that the
above functor be faithful.  
\end{description}
With full completeness, one has the tightest possible connection between
syntax and semantics.  We are not aware of any previously published
results of this type; however, the idea is related to representation
theorems in category theory~\cite{FS91}; to full abstraction theorems in
programming language semantics~\cite{Mil75,Plo77}; to studies of parametric
polymorphism~\cite{BFSS89,HRR89}; and to the completeness conjecture
in~\cite{GirC91}.

We now make a first statement in broad terms of our results.  We have
refined Blass' game semantics for Linear Logic. 
This refinement is not a complication; on the contrary, it makes the
definitions smoother and more symmetric.  Thus, we get a categorical model
of the logic, while Blass does not.  Then, we prove a Full
Completeness Theorem for this semantics, with respect to MLL + MIX
(Multiplicative
Linear Logic plus the Mix Rule).  Recall that the MIX rule~\cite{Gir87} has
the form 
\[ \mytwotrans{\entails \Gamma \; \; \; \; \entails \Delta}{\entails
\Gamma,\Delta} \]
There is a notion of proof net for this logic: this uses the Danos/Regnier
criterion~\cite{DanReg89}, simply omitting the connectedness part.  Thus, a proof structure
will be a valid proof net for MLL + MIX just if, for every switching, the
corresponding graph is acyclic.  This criterion was studied by Fleury and
Retor\'e~\cite{FR91}, used by Blute in his work on coherence
theorems~\cite{Blu92}, and adapted by Lafont for his work on interaction
nets~\cite{Laf90}.

Now we can state our result in more precise terms.  

\begin{theorem}
Every proof net in MLL + MIX denotes a uniform, history independent winning
strategy for  Player in our game interpretation.  Conversely, every
such strategy is the denotation of a unique cut-free proof net.
\end{theorem}

Of course, we now have to explain uniform, history independent strategies.
Note that a formula in MLL + MIX is built from atomic formulas and the
binary connectives tensor and par.  Its denotation will then be a {\em
variable type}.  We construe this as a functor over a category of games and
embeddings, in the fashion of domain theoretic semantics of
polymorphism~\cite{Gir86,CGW87}.  (In fact, this interpretation of variable types
is part of our game theoretic semantics of polymorphism).  An element of
variable type, the denotation of a proof of $\Gamma(\vec{\alpha})$, where
$\vec{\alpha}$ enumerates the atoms occurring in $\Gamma$, will then be a
family of strategies $\{\sigma_{\vec{A}}\}$, one for each tuple of
games $\vec{A}$ instantiating $\vec{\alpha}$.  The uniformity of this
family is expressed by the condition that it is a natural transformation
$\sigma: F^- \rightarrow F^+$, where $F^-, \;F^+$ are functors derived from
$\Gamma$ as explained in Section~\ref{vartypes}.

A history independent strategy is one in which the player's move is a
function only of the last move of the opponent and not of the preceding
history of the play.  Thus such a strategy is induced by a partial function
on the set of moves in the game.  The interpretation of proofs in MLL + MIX
by strategies, when analysed in terms of these underlying functions on
moves, turns out to be very closely related to the Geometry of Interaction
interpretation~\cite{TGI,GI1,GI2}.

The contents of the reminder of this paper are as follows.
Section~\ref{Mllmix} reviews MLL + MIX.  Section~\ref{core}
describes our game semantics for MLL + MIX.  Section~\ref{proofs} is devoted
to the proof of the Full Completeness Theorem.
Section~\ref{Fullinterpretation} outlines how our semantics can be extended
to full Classical Linear Logic.  Section~\ref{relatedwork} makes
comparisons with related work.

\section{MLL+MIX}\label{Mllmix}

The formulas $A,\; B, \;C, \ldots$ of MLL + MIX are built up from
propositional atoms $\alpha,\; \beta,\;\gamma, \ldots$ and their linear
negations $\alpha^{\perp},\; \beta^{\perp},\;\gamma^{\perp}, \ldots$ by
tensor (\tensor) and par (\parc).  The sequent calculus presentation of
MLL + MIX is as follows.  

\begin{center}
\begin{tabular}{||l|c|c||}\hline &&\\
Identity Group & 
\begin{tabular}{c}
\mytwotrans{}{ \entails  \alpha^{\perp},  \alpha} \\
\\
Identity
\end{tabular}
& 
\begin{tabular}{c}
\mytwotrans{ \entails \Gamma,  A \;\; \;\;\;\; \entails 
\Delta,  A^{\perp}}{   \entails  \Gamma,\Delta} \\
\\
Cut
\end{tabular}
\\ && \\ \hline && \\ 
Structural Group&  
\begin{tabular}{c}
\mytwotrans{\entails \Gamma}{ \entails \sigma \Gamma} \\
\\
Exchange 
\end{tabular}
& 
\begin{tabular}{c}
\mytwotrans{\entails \Gamma \;\; \;\;\;\; \entails \Delta }{   \entails
\Gamma, \Delta} \\
\\
Mix 
\end{tabular}
\\ && \\ \hline && \\
Multiplicatives & 
\begin{tabular}{c}
\mytwotrans{\entails \Gamma, A \;\;\;\;\;\;  \entails
 \Delta,  B}{   \entails \Gamma,\Delta,  A \tensor B}  \\
\\
Tensor
\end{tabular}
&
\begin{tabular}{c}
\mytwotrans{ \entails \Gamma,  A,  B}{
  \entails \Gamma,  A \parc B}  \\ 
\\
Par
\end{tabular}
 \\
\hline   
\end{tabular}
\end{center}
We have restricted the Identity axioms to propositional atoms; this does not
affect provability.  

\subsection{An aside: Units}
Our presentation has not included the {\em units} $\tunit$ for Tensor and
$\perp$ for Par.  The rules for these, {\em together} with the nullary
version of MIX, would be as follows.

\begin{center}
\begin{tabular}{||c|c|c||}\hline
Tensor Unit & Par Unit & Mix0 \\ \hline && \\
\mytwotrans{}{\entails \tunit} & \mytwotrans{\Gamma}{\entails
\Gamma,\perp}&\mytwotrans{}{\entails } \\ \hline
\end{tabular}
\end{center}

In fact, in the presence of the units, MIX can equivalently be expressed
by declaring $\tunit = \perp$.  It is easily checked that MIX and MIX0 are
derivable from this, and conversely that $\entails \tunit,\tunit$ and $\entails
\perp, \perp$ are derivable from MIX and MIX0.  But with $\tunit = \perp$,
clearly any sequent will be equivalent to one in which the units do not
occur.  Thus, we prefer to omit the units from our system.  

\subsection{Proof nets for MLL+MIX}

Proof structures can be defined for MLL + MIX just as for
MLL~\cite{Gir87,DanReg89}.  Alternatively, since we only allow atomic
instances of identity axioms, we can define a proof structure to be a pair
$(\Gamma, \phi)$, where $\Gamma $ is a sequent and $\phi$ is a fixpoint
free involution on the set of occurrences of literals in $\Gamma$, such
that, if $o$ is an occurrence of $l$, $\phi(o)$ is an occurrence of
$l^{\perp}$.  Thus, $\phi$ specifies the axiom links of the proof
structure; all the other information is already conveyed by $\Gamma$.  

A {\em switching} $S$ for a proof structure $(\Gamma, \phi)$ is an
assignment of L or R to each occurrence of \parc\ in $\Gamma$.  We then
obtain a graph $G(\Gamma,\phi,S)$ from the formation trees of the formulas
of $\Gamma$, together with the axiom links specified by $\phi$, with
unswitched arcs as specified by $S$ deleted.

{\flushleft {\bf Example: }}
\begin{eqnarray*}
\Gamma &=& \alpha_1^{\perp} \parc_0 \alpha_2^{\perp}, \alpha_3 \tensor
\alpha_4  \;\;\;\; \; \mbox{(subscripts are used to label occurrences)} \\
\phi &=& 1 \leftrightarrow 4, 2 \leftrightarrow 3 \\
S &=& 0 \mapsto {\mbox L}
\end{eqnarray*}

Then $G(\Gamma,\phi, S)$ is:

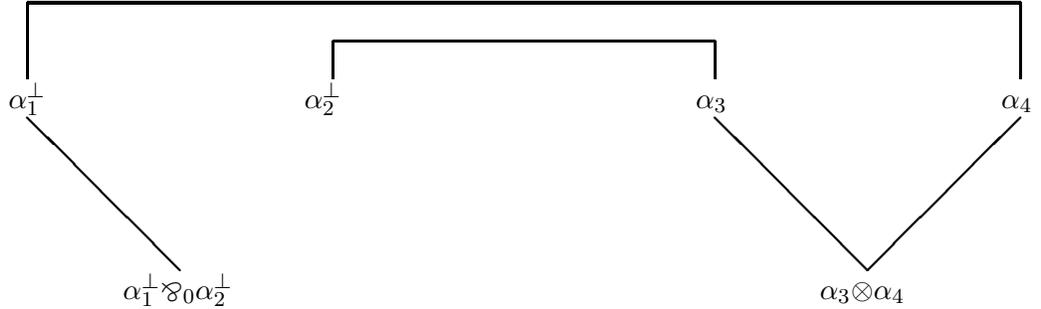
\setlength{\unitlength}{0.0100in}
\begin{picture}(574,170)(105,435)
\thicklines
\put(280,560){\line( 0, 1){ 20}}
\put(280,580){\line( 1, 0){200}}
\put(480,580){\line( 0,-1){ 20}}
\put(120,560){\line( 0, 1){ 40}}
\put(120,600){\line( 1, 0){520}}
\put(640,600){\line( 0,-1){ 40}}
\put(200,460){\line(-1, 1){ 80}}
\put(560,460){\line( 1, 1){ 80}}
\put(560,460){\line(-1, 1){ 80}}
\put (530,445) {\makebox(0,0) [lb] {\raisebox{0pt}[0pt][0pt]{ $\alpha_3 \tensor \alpha_4$}}}
\put (165,445) {\makebox(0,0) [lb] {\raisebox{0pt}[0pt][0pt]{ $\alpha^{\perp}_1 \parc_0 \alpha^{\perp}_2$}}}
\put (625,545) {\makebox(0,0) [lb] {\raisebox{0pt}[0pt][0pt]{ $\alpha_4$}}}
\put (465,545) {\makebox(0,0) [lb] {\raisebox{0pt}[0pt][0pt]{ $\alpha_3$}}}
\put (260,545) {\makebox(0,0) [lb] {\raisebox{0pt}[0pt][0pt]{ $\alpha^{\perp}_2$}}}
\put (105,545) {\makebox(0,0) [lb] {\raisebox{0pt}[0pt][0pt]{ $\alpha^{\perp}_1$}}}
\end{picture}

\begin{df}
A (cut-free) proof net for MLL+MIX is a proof structure $(\Gamma,
\phi)$ such that, for all switchings $S$, $G(\Gamma,\phi, S)$ is acyclic.
\end{df}

Fleury and Retor\'{e}~\cite{FR91} make a detailed study of this
criterion, which is of course just a modification of the Danos-Regnier
criterion~\cite{DanReg89}, to accomodate the MIX rule by dropping the
connectedness condition.  We can regard proof nets as the canonical
representations of (cut-free) proofs in MLL + MIX.

\section{The Game Semantics}\label{core}
\subsection{Basic Notions on Games}\label{basicintuitions}
This section describes the basic notions of Game and Strategy and relates
these ideas to Domain Theory and Processes.

We begin by fixing some notation.  If $X$ is a set, we write $X^{\star}$
for the set of finite sequences (words, strings) on $X$ and $X^{\omega}$
for the set of infinite sequences.  If $f: X \rightarrow Y$, then
$f^{\star}: X^{\star} \rightarrow Y^{\star}$ is the unique monoid
homomorphism extending $f$.  We write $|s|$ for the length of a
finite sequence.  If $Y \subseteq X$ and $s \in X^{\star}$, we write $s
{\upharpoonright} Y$ for the result of deleting all occurrences of symbols
not in $Y$ from $s$.  If $a \in X$ and $s \in X^{\star}$, we write $a \cdot
s$ ($s \cdot a$) for the result of prefixing (postfixing) $s$ with $a$.  We
write $ s \sqsubseteq t$ if $s$ is a prefix of $t$, {\em i.e.} for some $u$
$s u = t$.  We always consider sequences under this prefix ordering and use
order-theoretic notions~\cite{DP90} without further comment.

\subsubsection{Games}
The games we consider are between Player and Opponent.  A {\em play} or {\em
run} of the game consists of an alternating sequence of moves, which may be
finite or infinite.  Each play has a determinate outcome; one player wins
and the other loses.  Our plays are always with Opponent to move first.

\begin{df}
A game is a structure $A = (M_A,\lambda_A,P_A,W_A)$, where
\begin{itemize}
\item $M_A$ is the set of moves.
\item $\lambda_A: M_A \rightarrow \{ P,O \}$ is the labelling function to
indicate if a move is by Player or Opponent.  We write $M_A^+ =
\lambda_A^{-1}(\{P \}), \; M_A^- = \lambda_A^{-1}(\{O \})$ and
$\overline{P} = O, \; \overline{O} = P$.
\item Let $M_A^{\circledast}$ be the set of all alternately-labelled finite
sequences of moves, {\em i.e.} 
\[ M_A^{\circledast} = \{ s \in M_A^{\star} \mid \; (\forall i: 1 \leq i
< |s| )\; [ \lambda_A(s_{i+1}) = \overline{\lambda_A(s_{i})} ] \} \]
Then $P_A$, the set of valid {\em positions} of the game, is a non-empty
prefix closed subset of $M_A^{\circledast}$.
\item Let $P_A^{\infty}$ be the set of all infinite sequences of moves, all
of whose finite prefixes are in $P_A$.  $W_A$ is a subset of
$P_A^{\infty}$, indicating which infinite plays are won by Player.
\end{itemize}
\end{df}
{\bf An Important Remark: }
Note that $P_A$ may contain positions in which the opening move is by
Player, even though all {\em plays} in $A$ must be started by Opponent.
This becomes significant when games are combined, {\em e.g.} with tensor.
Sections~\ref{Fullinterpretation} and~\ref{relatedwork} discuss 
this point in detail.

\subsubsection{Strategies}
A {\em strategy} for Player (with Opponent to start) in $A$ is usually
defined to be a partial function from positions (with Player to move) to moves
(by Player).  We prefer the following definition, which leads to a more
elegant treatment of composition. 

\begin{df}
A {\em strategy} is a non-empty prefix closed subset $\sigma \subseteq P_A$
satisfying
\begin{description}
\item[(s1)] $ a \cdot s \in \sigma \; \impl\ \lambda_A(a) = O$.  
\item[(s2)] If $s \cdot a, \; s \cdot b \in \sigma$, Player to move at $s$,
then $a=b$.
\item[(s3)] If $s \in \sigma$, Opponent to move at $s$, $s \cdot a \in
P_A$, then $s \cdot a \in \sigma$.  
\end{description}
\end{df}
Of these conditions, the first incorporates the convention that Opponent is
to start; and the second enforces that strategies are {\em deterministic}.
Note that any strategy $\sigma$ does indeed determine a partial function
$\hat{\sigma}$ on positions with Player to move.

We can readily define the notion of a strategy for Opponent (with Opponent
to start) in $A$, by interchanging Player and Opponent in
conditions {\bf (s2)} and {\bf (s3)}.  Such a strategy is called a {\em
counter-strategy}.  Given a strategy $\sigma$ and a counter-strategy
$\tau$, we can define the play that results when Player follows $\sigma$
and Opponent follows $\tau$:
\[ \langle \sigma \mid \tau \rangle \ = \lub \; (\sigma \cap \tau) \]
Here $\sigma \cap \tau$ is an ideal of the poset $P_A$, in fact
a down-closed chain.  Its join $s$, taken in the directed completion of
$P_A$, $P_A \cup P_A^{\infty}$, is a finite or infinite play.  In the
former case, the player who is to play at $s$ loses; in the latter case,
Player wins if and only if $s \in W_A$.
A strategy is {\em winning} if it beats all counter-strategies.

\subsubsection{Games and Domain theory}
The following table draws an analogy between games and Domain theory.
\begin{center}
\begin{tabular}{||l|l||}\hline 
Game & Information System \\ \hline
Strategy & Domain Element \\ \hline
Winning Strategy & Total Element \\ \hline
\end{tabular}
\end{center}

\subsubsection{Games and Processes}
The following table draws a much richer analogy between games and
concurrent processes.  
\begin{center}
\begin{tabular}{||l|l||}\hline 
Game & Process Specification \\ \hline
Moves & Alphabet or Sort of actions  \\ \hline
Player  & System  \\ \hline
Opponent   & Environment  \\ \hline
$P_A$ & Safety specification \\ \hline
$W_A$ & Liveness specification \\ \hline 
Strategy & Process \\ \hline
Strategy in $A$ & \hspace{-6pt}\begin{tabular}{l}
Process satisfying safety specification \\
``Partial correctness'' 
\end{tabular} \\ \hline 
Winning Strategy & \hspace{-6pt}\begin{tabular}{l}
Deadlock-free process satisfying liveness specification \\
``Total correctness'' 
\end{tabular} \\ \hline
\end{tabular}
\end{center}

\subsection{The Game interpretation of the Multiplicatives}\label{definitions}

\subsubsection*{Linear Negation}
\[ A^{\perp} = (M_A,\overline{\lambda_A}, P_A,P_A^{\infty} \setminus W_A)
\]
where $\overline{\lambda_A}(a) =  \overline{\lambda_A(a)}$.  Clearly
$A^{\perp\perp} = A$.

\subsubsection*{Tensor}

The game $A \tensor B$ is defined as follows. 
\begin{itemize}
\item $M_{A \tensor B} = M_A + M_B$, the disjoint union of the two move
sets.
\item $\lambda_{A \tensor B} = [\lambda_A,\lambda_B]$, the source tupling. 
\item $P_{A \tensor B}$ is the set of all alternately labelled finite sequences
of moves such
that:
\begin{enumerate}
\item The restriction to the moves in $M_A$ (resp. $M_B$) is in
$P_A$ (resp. $P_B$)
\item If two successive moves are in different components, ({\em i.e.} one
is in $A$ and the other is in $B$), it is the Opponent who has switched
components. 
\end{enumerate}

\item $W_{A \tensor B}$ is the set of infinite plays of the game, such that
the restriction to each component is either finite or is a win for Player
in that component.
\end{itemize}
The tensor unit is given by
\[ \tunit = (\varnothing,\varnothing,\{\epsilon \}, \varnothing) \]
Note that $ {\perp} = \tunit^{\perp} = \tunit$.

\subsubsection*{Other Connectives}
The other multiplicative connectives can be defined from Tensor and
Linear negation:  
\begin{eqnarray*}
A \parc B &=& (A^{\perp} \tensor B^{\perp})^{\perp} \\
A \linimpl B &=& A^{\perp} \parc B 
\end{eqnarray*}

\subsubsection*{Comment on the definitions}
Note that positions in $A$ with first move by Player can indeed be
significant for plays in $A^{\perp}, A \tensor B$ etc.  This will be more
fully discussed in relation to Blass' definitions in
Section~\ref{relatedwork}.  The main point that we wish to make here is that
there are clear intuitions behind our definition of $P_{A \tensor B}$
(and similarly of $P_{A \parc B}, \; P_{A \linimpl B}$).

The first condition on $P_{A \tensor B}$ says that a play in $A \tensor B$
consists of (an interleaved representation of) concurrent plays in $A$ and
$B$.  (Compare this with the definition of composition without
communication in the trace model of CSP~\cite{Hoa85}).  The second
condition, that Player must move in the same component in which Opponent last
moved, while Opponent is free to switch components, reflects the
fundamental meaning of, and difference between Tensor and Par.
Tensor is {\em disjoint} concurrency; Par is {\em connected} concurrency.
That is, Tensor combines two processes in parallel with no flow of
information between them; while Par allows flow of information.  (More
precisely, in MLL flow is {\em required} for Par; this is the content of
the connectedness part of the proof-net criterion.  In MLL + MIX, flow
is {\em permitted} but not obligatory, so that Tensor becomes a special
case of Par.)  These constraints on the flow of information are reflected
in game-theoretic terms as follows.  The Player for Tensor (or Opponent for
Par) must respond in the component in which his adversary moved; while
Opponent for Tensor (or Player for Par) is allowed to use the moves of his
adversary in one component to influence his play in the other component.
In this way we get the chess game strategy by which I can defeat
Karpov or Kasparov if I play against them in the following configuration\footnote{This example is taken from~\cite{LafStr91}, but the same idea can
be found in~\cite{Con76}.}: 
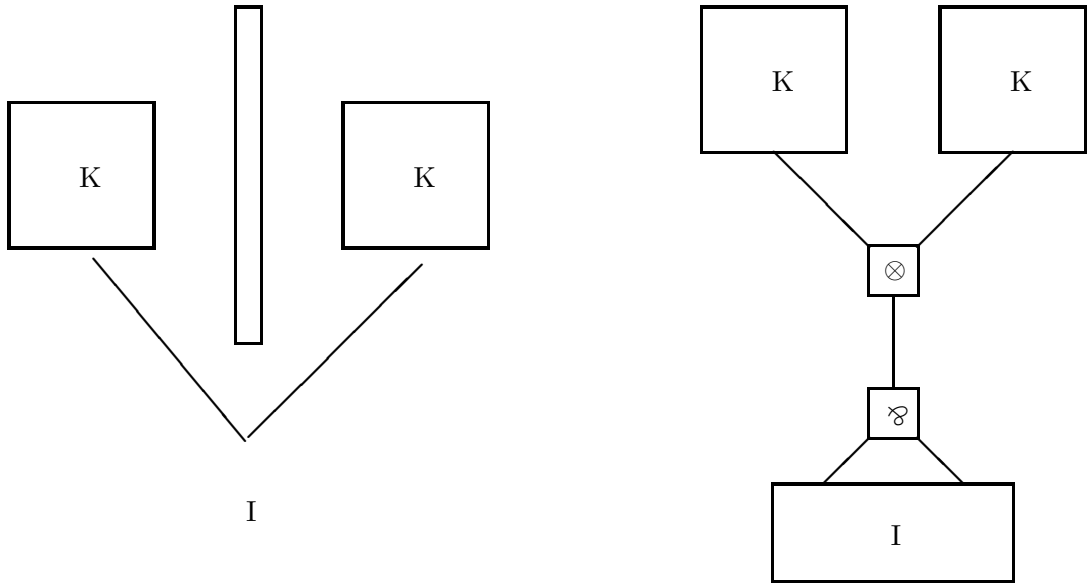
\begin{center}
\setlength{\unitlength}{0.0125in}
\begin{picture}(450,255)(80,450)
\thicklines
\put( 80,600){\framebox(60,60){}}
\put(220,600){\framebox(60,60){}}
\put(115,595){\line( 5,-6){ 63.525}}
\put(180,520){\line( 1, 1){ 72.500}}
\put(175,560){\framebox(10,140){}}
\put(470,640){\framebox(60,60){}}
\put(370,640){\framebox(60,60){}}
\put(440,580){\framebox(20,20){}}
\put(440,520){\framebox(20,20){}}
\put(440,520){\line(-1,-1){ 20}}
\put(460,520){\line( 1,-1){ 20}}
\put(400,460){\framebox(100,40){}}
\put(450,580){\line( 0,-1){ 40}}
\put(400,640){\line( 1,-1){ 40}}
\put(460,600){\line( 1, 1){ 40}}
\put (105,625) {\makebox(0,0) [lb] {\raisebox{0pt}[0pt][0pt]{ K}}}
\put (245,625) {\makebox(0,0) [lb] {\raisebox{0pt}[0pt][0pt]{ K}}}
\put (175,485) {\makebox(0,0) [lb] {\raisebox{0pt}[0pt][0pt]{ I}}}
\put (495,665) {\makebox(0,0) [lb] {\raisebox{0pt}[0pt][0pt]{ K}}}
\put (445,475) {\makebox(0,0) [lb] {\raisebox{0pt}[0pt][0pt]{ I}}}
\put (395,665) {\makebox(0,0) [lb] {\raisebox{0pt}[0pt][0pt]{ K}}}
\put (442,587) {\makebox(0,0) [lb] {\raisebox{0pt}[0pt][0pt]{ $\tensor$}}}
\put (440,526) {\makebox(0,0) [lb] {\raisebox{0pt}[0pt][0pt]{ $\hspace{1mm}\parc$}}}
\end{picture}
\end{center}
and I play white in one game and black in the other.  (The vertical
rectangle represents a screen between Karpov and Kasparov that prevents
each from seeing the other's game board, while I can see both games).  This
``copy-cat'' strategy is the game-theoretic content of the Identity axiom
$\entails A^{\perp},A$ (or equivalently $\entails A^{\perp} \parc A$).

These ideas can also be related to the trip condition for proof
nets~\cite{Gir87}: the difference between Tensor and Par is expressed thus in
terms of the trip condition (\cite{Gir87} Introduction, Section III.4.3):
\begin{itemize}
\item ``In the case of  \tensor\, there is no cooperation: if we start with 
\que{A}, then we come back through \ans{A} before entering \que{B} after
which we come back through \ans{B}.  
\item in the case of \parc\, there is cooperation: if we start again with
\que{A}, then we are expected through \ans{B}, from which we go to \que{B}
and eventually come back through \ans{A}.'' 
\end{itemize}

Thus we get the following possible transitions in trips:
\begin{description}
\item[$A \tensor B$: ] $\que{A}\ans{A}  \que{B}\ans{B}$ or $\que{B}\ans{B} \; \que{A}\ans{A}$ 
\item[$A \parc B$: ] $\que{A}\ans{B}  \que{B}\ans{A}$ or $\que{B}\ans{A}
\que{A}\ans{B}$ 
\end{description}

If we correlate ``questions'', in the terminology of~\cite{Gir87}, with
moves by Opponent and ``answers'' with moves by Player, this says exactly that
only Opponent (Player) may switch between components in a Tensor (Par)
game.

\subsection{The Category of Games}\label{gamecategory}
We build a category \Games\ with games as objects and  winning
strategies as morphisms.  The objects of \Games\ are games; the morphisms
$\sigma: A \rightarrow B$ are the winning strategies in $A
\linimpl B = A^{\perp} \parc B$. 

The composition of strategies can be defined elegantly
in terms of the set representation.  Firstly, a preliminary definition.
Given a sequence of games $A_1, \ldots, 
A_n$, we define ${\cal L}(A_1,\ldots A_n)$, the {\em local} strings on
$A_1, \ldots, A_n$, to be the set of all $s \in (M_{A_1}+
\cdots+M_{A_n})^{\star}$ such that, for all $i$ with $1 \leqslant i < |s|$,
$s_i \in M_{A_j}$ and $s_{i+1} \in M_{A_k}$ implies that $j$ is adjacent
to $k$, {\em i.e.} $|j -k| \leqslant 1$.  Now, given $\sigma: A \rightarrow B, \;
\tau: B \rightarrow C$, define
\[ \sigma; \tau =  \{ s {\upharpoonright} A,C \mid \; s \in {\cal
L}(A,B,C), \; s {\upharpoonright} A,B \in \sigma, \; s {\upharpoonright}
B,C \in \tau \} \] 
Here, $s {\upharpoonright} X,Y$ means the
result of deleting all moves in $s$ not in $M_X$ or $M_Y$.  Note that this
definition clearly exhibits the ``Cut = Parallel Composition + Hiding''
paradigm proposed by the first author~\cite{AbrP91} as the correct
computational interpretation of Cut in Classical Linear Logic, with respect
to the CSP-style trace semantics for parallel composition and
hiding~\cite{Hoa85}.  What makes the game semantics so much richer than
trace semantics is the explicit representation of the environment as the
Opponent.

\begin{proposition}\label{compstrategy}
If $\sigma: A \rightarrow B, \; \tau: B \rightarrow C$ are winning
strategies, so is $\sigma;\tau$.
\end{proposition}
\pr\ Let $S = \{ s \in {\cal L}(A,B,C) \mid s {\upharpoonright} A,B \in
\sigma, \; s {\upharpoonright} B,C \in \tau \}$ so that $\sigma;\tau = \{ s
{\upharpoonright} A,C \mid \; s \in S \} $.  Firstly, note that
$\sigma;\tau$ is non-empty and prefix closed because $S$ is.  

Since $s \in S$ implies $s {\upharpoonright} A,B \in \sigma$, $(s
{\upharpoonright} A,C){\upharpoonright} A  = ( s {\upharpoonright}
A,B){\upharpoonright} A \in P_A$ and similarly, $(s
{\upharpoonright} A,C){\upharpoonright} C \in P_C$.

Now, suppose $s {\upharpoonright} A,C = t \cdot a \cdot c \in \sigma; \tau$
with $ s \in S, \; a \in M_A, \; c \in M_C$.  Since $ s \in {\cal
L}(A,B,C)$, we must have $s = s' \cdot a \cdot b_1\cdot \ldots \cdot b_k
\cdot c$, for some $b_1,\ldots,b_k \in M_B$ with $ k \geq 1$.  Moreover,
\begin{eqnarray*}
(s' {\upharpoonright} A,B) \cdot a \cdot b_1\cdot \ldots \cdot b_k & \in&
\sigma \\
(s' {\upharpoonright} B,C) \cdot  b_1\cdot \ldots \cdot b_k \cdot c & \in&
\tau
\end{eqnarray*}
Hence, $a$ must be an $O$-move and $c$ must be a $P$-move.  A symmetric
argument applies when $t \cdot c \cdot a \in \sigma; \tau$.  We have shown
that $ \sigma; \tau \subseteq P_{A \linimpl C}$.

Next, note that if $s \in S$, $s$ cannot start with a move in $B$ since
this would violate {\bf (s1)} either for $s {\upharpoonright} A,B \in
\sigma$, or for $s {\upharpoonright} B,C \in \tau$.  If $s = a \cdot s'$
with $a \in M_A$, then $a \cdot (s'{\upharpoonright} A,B) \in
\sigma $, so $a$ is an $O$-move  by {\bf (s1)} applied to $ \sigma$; and
similarly if $ s= c \cdot s'$ with $c \in M_C$.  Thus, $\sigma; \tau$
satisfies {\bf (s1)}.  

Given $t \in \sigma;\tau$ we say that $s$ {\em covers} $t$ if
\begin{itemize}
\item $s \in {\cal L}(A,B,C)$
\item $s {\upharpoonright} A,C = s$
\item $s {\upharpoonright} A,B \in \sigma, \; s {\upharpoonright} B,C \in
\tau$
\end{itemize}
We claim that for each $t \in \sigma;\tau$ there is a {\em least} $s$
covering $t$; we write $s \succ t$ in this case.  Moreover, we claim that
if $t \in \sigma; \tau$ with Opponent to move at $t$, then for any $d$ such
that $t \cdot d \in P_{A \linimpl C}$, there is a unique $e$ such that $t
\cdot d \cdot e \in \sigma; \tau$.  We will prove these claims by
simultaneous induction on $|t|$.
\begin{itemize}
\item $\epsilon \succ \epsilon$
\item If $t =t'\cdot d$, where $d$ is an $O$-move, then by induction we
have $s' \succ t'$, and then $s = s'\cdot d \succ t' \cdot d =t$.  Note
that this is well defined: since $t' \cdot d$ is in $P_{A \linimpl C}$,
either $t' =\epsilon$ or $d$ is in the same component as the previous
$P$-move.  By minimality of $s'$, either $s'= \epsilon$ or $s' = s'' \cdot
e $, where $e$ is the previous $P$-move in $t$.  In either case, $s
{\upharpoonright} A,B \in \sigma, \; s {\upharpoonright} B,C \in \tau$ as
required.  

\item  If $t =t' \cdot d$, where $d$ is an $O$-move, then by
induction hypothesis, we have $s= s' \cdot d \succ t$.  Suppose $d
 \in M_A$ (the case of $d \in M_C$ is symmetrical). 

Since $\sigma$ is a winning strategy in $ A \linimpl B$, it has a
unique response $e$ to $(s {\upharpoonright} A,B) \cdot d$, which is either
$ e= a'\in A$, or $ e = b_1 \in B$.  Moreover, $e$ is the unique move such
that $s' \cdot d \cdot e \in S$, by the requirements that $e$ is in $A$ or $B$
and that $(s' \cdot d \cdot e) {\upharpoonright} A,B \in \sigma$.  If $e =
b_1$, then $b_1$ is an $O$-move in $B^{\perp}$, and since $ \tau$ is a
winning strategy in $B \linimpl C$, it has a unique response to $( s \cdot
d \cdot b_1) {\upharpoonright} B,C$, which will be either $b_2 \in B$ or
$c' \in C$.  Continuing in this way, we obtain a uniquely determined
sequence of extensions of $s$ in $S$.  Either this sequence culminates in
$s \cdot d \cdot b_1 \cdot \ldots \cdot b_k \cdot e$, where $e$ lies in $A$
or $C$, or the sequence of ``internal'' moves in $B$ is infinite.  We claim
that the latter situation cannot in fact apply; for if it did, we would
have infinite plays $u = (s \cdot d \cdot b_1 \cdot b_2 \cdots )
{\upharpoonright} A,B$ in $A \linimpl B$ following $\sigma$ and $v = (s
\cdot d \cdot b_1 \cdot b_2 \cdots ) {\upharpoonright} B,C$ in $B
\linimpl C$ following $\tau$.  Since $u {\upharpoonright} A$ and $v
{\upharpoonright} C$ are finite, and $u {\upharpoonright B} = v
{\upharpoonright} B^{\perp}$, Player must lose in one of these plays,
contradicting the hypothesis that $\sigma$ and $\tau$ are both winning.  It
is clear that $s \cdot d \cdot b_1\cdot \ldots \cdot b_k \cdot e \succ t
\cdot d \cdot e$.   
\end{itemize}
Thus $\sigma; \tau$ satisfies {\bf (s2)}, and moreover has a well defined
response at all positions with Player to move.  It remains to be shown that
if Player follows $\sigma;\tau$ he wins all infinite plays.  Let $s$ be
such a play; we must show that if $s {\upharpoonright} A \in W_A$ or $s
{\upharpoonright} A$ is finite, then $s {\upharpoonright} C \in W_C$.  Let
$\{ s_k \}$ be the increasing sequence of finite prefixes of $s$.  Let $\{
t_k \}$ be the corresponding increasing sequence where $t_k \succ s_k$.  Let $t
= \lub t_k$.  Then $t {\upharpoonright} A,B$ is an infinite play following
$\sigma$ and $t {\upharpoonright} B,C$ is an infinite play following
$\tau$.  If $s {\upharpoonright} A = t {\upharpoonright} A \in W_A$, then
since $\sigma$ is winning, $t {\upharpoonright} B \in W_B$; and then since
$\tau$ is winning, $t {\upharpoonright} C = s {\upharpoonright} C \in W_C$,
as required.  \epr\

Note that part of what we proved is that when two winning strategies are
composed, we cannot get infinite ``chattering'' ({\em i.e.} internal
communication) in the terminology of CSP~\cite{Hoa85}. 

\begin{proposition}
\Games\ is a category.
\end{proposition}
\pr\ We define the identity morphism ${\tt id}_A: A \rightarrow A$ as 
\[ {\tt id}_A = \{ s \in P_{A \linimpl A} \mid \mbox{ $s$ begins with an
$O$-move}, \;(\forall t \sqsubseteq s) \; ( |t|
\; \mbox{even} \; \impl\  t {\upharpoonright} A = t {\upharpoonright} A^{\perp}
) \}  \]
In process terms, this is a bidirectional one place
buffer~\cite{AbrP91}.  In game terms, this is the copy-cat
strategy discussed previously.  

Next, we prove associativity.  Given $\sigma: A \rightarrow B,\; \tau: B
\rightarrow C, \; \upsilon : C \rightarrow D$, we will show that 
$ (\sigma;\tau);\upsilon = S $,
where 
\[ S = \{ t{\upharpoonright} A,D \mid t \in {\cal L}(A,B,C,D), \; t
{\upharpoonright} A,B \in \sigma, \;t {\upharpoonright} B,C 
\in \tau,\; t {\upharpoonright} C,D \in \upsilon \} \]
A symmetrical argument shows that $ \sigma;(\tau;\upsilon) = S $, whence we
get the required result.

The inclusion $S \subseteq (\sigma;\tau);\upsilon $ is straightforward. Write 
\begin{eqnarray*}
(\sigma;\tau); \upsilon &=& \{ s {\upharpoonright} A,D \mid s \in {\cal L}(A,C,D), \; s {\upharpoonright} C,D \in \upsilon, \\
&& \;\;\;\;\;\;\;\;\;\; (\exists t \in {\cal L}(A,B,C))\; [  t {\upharpoonright} A,B \in \sigma, \; t
{\upharpoonright} B,C \in \tau , \; t {\upharpoonright} A,C = s
{\upharpoonright} A,C ] \} 
\end{eqnarray*}
Given $u {\upharpoonright} A,D \in S$, $ u {\upharpoonright} A,B,C $
witnesses that $ u {\upharpoonright} A,C \in \sigma; \tau$, while $ u
{\upharpoonright} C,D \in \upsilon$ by assumption.  Hence, $u {\upharpoonright}
A,D \in (\sigma;\tau);\upsilon$.  

For the converse, a witness $t$ such that $t {\upharpoonright} A,D \in S$
may be constructed from $s \in (\sigma;\tau); \upsilon$ by the same argument used to construct $t \succ
s$ in Proposition~\ref{compstrategy}.  \epr\

\subsection{History-free strategies}
We will be interested in a restricted class of strategies, the history-free
(or history independent, or history insensitive) ones. A strategy for Player
is history-free if there is some partial function $f: M_A^-
\rightarrow M_A^+$, such that at any position $s\cdot a$, with Player to
move, 
\[ \hat{\sigma}(s\cdot a) = \left\{ \begin{array}{l}
                                f(a), \;\;\;\;\;\;\;\;\;\; f(a) \; \mbox{defined and} \;  s
\cdot a \cdot f(a) \in P_A \\
\mbox{undefined}, \; \mbox{otherwise }
\end{array}
\right. \]
Clearly, in this case, there is a {\em least} partial function inducing
$\sigma$; we write $\sigma = \sigma_f$, always meaning this least $f$.  It
is important to note that the category \Games\ described in
subsection~\ref{gamecategory} also forms a model of MLL + MIX.  However, to
obtain a precise correspondence with the logic, we will focus our attention
on the sub-category $\Games_{\tt hf}$ of history-free strategies.

A history-free strategy $\sigma = \sigma_f$ is uniquely determined by
the underlying function $f$ on moves.  In particular, all the morphisms
witnessing the $\star$-autonomous structure in $\Games_{\tt hf}$, or
equivalently the interpretations of proofs in MLL + MIX~\cite{Seeley}, can be defined
directly in terms of these functions.  When we do so, we find that the
interpretation coincides exactly with the Geometry of Interaction
interpretation~\cite{TGI,GI1,GI2}.  More precisely, it corresponds to a
reformulation of the Geometry of Interaction, due to the present authors,
in a typed version based on sets and partial functions, in the same spirit
as the $\Gi(\category{C})$ construction of~\cite{Abr92}.

\subsubsection{Games and the Geometry of Interaction}\label{gamesgoi}

As a first illustration, we consider composition again.  Say we have
$\sigma_f: A \rightarrow B, \; \tau_g: B\rightarrow C$.  We want to find
$h$ such that $\sigma_f; \tau_g = (\sigma;\tau)_h$.  We shall compute $h$
by the ``execution formula''~\cite{TGI,GI1,GI2}, cut down to its actual
content, which is adequately described in terms of sets and partial
functions.  Before giving the formal definition, let us explain the idea,
which is rather simple.  We want to hook the strategies up so that Player's
moves in $B$ under $\sigma$ get turned into Opponent's moves in $B^{\perp}$
for $\tau$, and vice versa.  Consider the following picture:
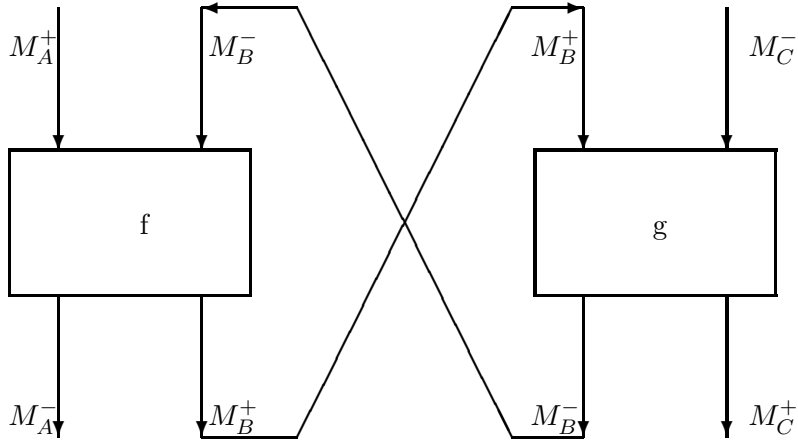
\begin{center}
\setlength{\unitlength}{0.0125in}
\begin{picture}(361,195)(135,550)
\thicklines
\put(350,560){\line( 1, 0){ 30}}
\put(220,560){\line( 1, 0){ 40}}
\put(350,740){\vector( 1, 0){ 30}}
\put(260,740){\vector(-1, 0){ 40}}
\put(260,560){\line( 1, 2){ 90}}
\put(260,740){\line( 1,-2){ 90}}
\put(440,620){\vector( 0,-1){ 60}}
\put(440,740){\vector( 0,-1){ 60}}
\put(380,620){\vector( 0,-1){ 60}}
\put(380,740){\vector( 0,-1){ 60}}
\put(220,620){\vector( 0,-1){ 60}}
\put(160,620){\vector( 0,-1){ 60}}
\put(220,740){\vector( 0,-1){ 60}}
\put(160,740){\vector( 0,-1){ 60}}
\put(360,620){\framebox(100,60){}}
\put(140,620){\framebox(100,60){}}
\put (445,565) {\makebox(0,0) [lb] {\raisebox{0pt}[0pt][0pt]{ $M^+_C$}}}
\put (445,720) {\makebox(0,0) [lb] {\raisebox{0pt}[0pt][0pt]{ $M^-_C$}}}
\put (360,565) {\makebox(0,0) [lb] {\raisebox{0pt}[0pt][0pt]{ $\hspace{-2mm}M^-_B$}}}
\put (360,720) {\makebox(0,0) [lb] {\raisebox{0pt}[0pt][0pt]{ $\hspace{-2mm}M^+_B$}}}
\put (225,565) {\makebox(0,0) [lb] {\raisebox{0pt}[0pt][0pt]{ $\hspace{-2mm}M^+_B$}}}
\put (225,720) {\makebox(0,0) [lb] {\raisebox{0pt}[0pt][0pt]{ $\hspace{-2mm}M^-_B$}}}
\put (135,565) {\makebox(0,0) [lb] {\raisebox{0pt}[0pt][0pt]{ $M^-_A$}}}
\put (135,720) {\makebox(0,0) [lb] {\raisebox{0pt}[0pt][0pt]{ $M^+_A$}}}
\put (405,645) {\makebox(0,0) [lb] {\raisebox{0pt}[0pt][0pt]{ g}}}
\put (190,645) {\makebox(0,0) [lb] {\raisebox{0pt}[0pt][0pt]{ f}}}
\end{picture}
\end{center}
Assume that the Opponent starts in $A$.  There are two
possible cases:
\begin{itemize}
\item  The move is mapped by $f$ to a response in $A$: In this case,
this is the response of the function $h$.
\item The move is mapped by $f$ to a response in $B$.  In this case, this
response is interpreted as a move of the Opponent in $B^{\perp}$ and fed as
input to $g$.  In turn, if $g$ responds in $C$, this is the response of the
function $h$.  Otherwise, if $g$ responds in $B^{\perp}$, this is fed back
to $f$.  In this way, we get an internal dialogue between the strategies
$f$ and $g$; this dialogue {\em cannot be infinite}, because $\sigma, \;
\tau$ are both {\em winning} strategies.
\end{itemize}
Thus, ``termination of Cut-elimination'', or nilpotency in
terms of the Geometry of Interaction, corresponds to ``no infinite internal
chattering''  in process-algebra terms.  

It remains to give a formula for computing $h$ according to these
ideas.  This is the execution formula:
\[ h = \bigvee_{k \in \omega} m_k \]
The join in the definition of $h$ can be interpreted concretely as union of
graphs.  It is well-defined because it is being applied to a family of
partial functions with pairwise disjoint domains of definition.  The
functions $m_k: M_A^+ + M_C^- \rightharpoonup M_A^- + M_C^+ $ are defined by
\[ m_k = \pi^{\star} \circ (( f + g) \circ \mu)^{k}  \circ (
f + g) \circ \pi \] The idea is that $m_k$ is the function which, when
defined, feeds an input from $M_A^+$ or $M^-_C$ exactly $k$ times around
the channels of the internal feedback loop and then exits from $M^-_A$ or
$M^+_C$.  The retraction
\[ \pi: M_A
+ M_C \lhd M_A +M_B+M_B+M_C : \pi^{\star} \]
is defined by
\[ 
\pi^{\star} = [{\tt inl},0,0,{\tt inr}] \;\;\;\;\;
\pi = [{\tt in}_1,{\tt in}_4] 
\]
and the ``message exchange'' function 
$\mu: M_A^- +M_B^++M_B^-+M_C^+ \rightharpoonup M_A^+ +M_B^-+M_B^++M_C^-$ is defined by 
\[ \mu = 0 + [{\tt inr},{\tt inl}] + 0 \]
Here, $0$ is the everywhere undefined partial function.

\subsubsection{The Category of Games and History-free strategies}

We build a category $\Games_{\tt hf}$ with games as objects and history-free
winning strategies as morphisms.  The objects of $\Games_{\tt hf}$ are games;
the morphisms $\sigma: A \rightarrow B$ are the history-free winning
strategies in $A \linimpl B = A^{\perp} \parc B$.

\begin{proposition}\label{hfsubcat}
$\Games_{\tt hf}$ is a sub-category of \Games.  
\end{proposition} 
\pr\ Note that the identity morphism ${\tt id}_A: A \rightarrow A$ is
history-free.  Thus, it suffices to prove that $\Games_{\tt hf}$ is closed
under composition.

Let $\sigma_f: A \rightarrow B$ and $\sigma_g: B \rightarrow C$ be
history-free winning strategies.  Then, with notation as above, we need
to show that: $ \sigma_f; \sigma_g = \sigma_h $.  We show that for all $s$
with $|s|$ even 
\[ s  \in \sigma_h \ifof\ s \in \sigma_f; \sigma_g \]
We argue by induction on $|s|$.  The basis $s = \epsilon$ is clear.  

Now, suppose $s \cdot d \cdot e \in \sigma_f; \sigma_g$.  From the proof of
Proposition~\ref{compstrategy}, we have that there exists 
\[ t = s' \cdot d
\cdot b_1 \cdot b_2 \ldots \cdot b_k \cdot e  \succ s \cdot d \cdot e \]
Suppose, for example, that $d$ is in $A$, e is in $C$; then $(f(d) = b_1,
g(b_1) = b_2, \ldots, f(b_{k-1}) = b_k, g(b_k) =e$.  But then, $m_k(d) =
e$, so $h(d) = e$ and applying the induction hypothesis to $s$, $s \cdot d
\cdot e \in \sigma_h$.

For the converse, suppose $s \cdot d \cdot e \in \sigma_h$.  Then, for some
$k$, $m_k(d) = e$, {\em i.e.} (again considering for example, the case
where $d$ is in $A$ and $e$ is in $C$), $f(d) = b_1, g(b_1) = b_2, \ldots,
f(b_{k-1}) = b_k, g(b_k) =e$.  By induction hypothesis, $s \in
\sigma_f;\sigma_g$, so for some $t \succ s$, $t {\upharpoonright} A,B \in
\sigma_f, t {\upharpoonright} B,C \in
\sigma_g$.  But then
\begin{eqnarray*}
(t {\upharpoonright} A,B ) \cdot d \cdot b_1 \cdot \ldots \cdot b_k &\in&
\sigma_f \\
(t {\upharpoonright} B,C ) \cdot  b_1 \cdot \ldots \cdot b_k \cdot e &\in&
\sigma_g 
\end{eqnarray*}
and so $s\cdot d \cdot e = (t \cdot d \cdot b_1 \cdot \ldots \cdot b_k \cdot
e){\upharpoonright} A,C \in \sigma_f;\sigma_g   $.   \epr

\subsection{$\star$-autonomous categories of games}

\subsubsection{$\Games_{\tt hf}$ as a $\star$-autonomous category}\label{ghfstar}

We show that $\Games_{\tt hf}$ is a $\star$-autonomous
category, and thus yields an interpretation of the
formulas and proofs of MLL + MIX.  (For background,
see~\cite{Seeley,Barr91}).  We have
already defined the object part of the tensor
product $A \tensor B$, the linear negation $A^{\perp}$ and the tensor unit.

The action of tensor on morphisms is defined as follows.  If $\sigma_f: A
\rightarrow B, \; \tau_g: A' \rightarrow B'$, then $\sigma \tensor \tau: A
\tensor A' \rightarrow B \tensor B'$ is induced by
\begin{eqnarray*}
h &=& (M_A^+ + M_{A'}^+) + (M_B^- + M_{B'}^-) \cong (M_A^+ + M_{B}^-) + (
M_{A'}^++ M_{B'}^-) \\
&& \stackrel{f+g}{\rightarrow} (M_A^- + M_{B}^+)+ ( M_{A'}^-+ M_{B'}^+) \\
&& \cong (M_A^- + M_{A'}^-) + (M_B^+ + M_{B'}^+)
\end{eqnarray*}

The natural isomorphisms for associativity, commutativity and unit of the
tensor product are induced from those witnessing the symmetric monoidal
structure of coproduct (disjoint union) in {\bf Set}; say
${\tt assoc}$,  ${\tt symm}$, ${\tt unit}$.  For
example, the associativity of Tensor is given by $\sigma_h: (A \tensor B)
\tensor C
\cong A \tensor (B \tensor C)$, where 
\[ h: ((M^+_A + M^+_B) + M^+_C) + (M^-_A + (M^-_B + M^-_C)) \cong ((M^-_A
+ M^-_B) + M^-_C) + (M^+_A + (M^+_B + M^+_C)) \]
is the canonical isomorphism constructed from ${\tt assoc}$ and ${\tt
symm}$.  

Similarly, the application morphism ${\tt apply}: (A \linimpl B) \tensor A
\rightarrow B$ is induced by 
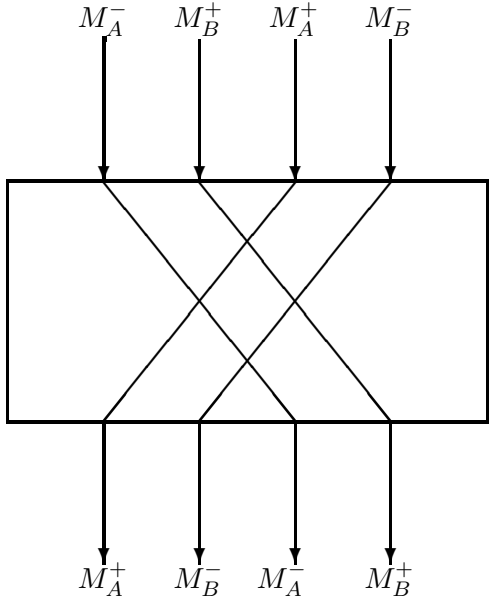
\[ (M^-_A + M^+_B) + M^+_A) + M^-_B \cong (M^+_A + M^-_B) + M^-_A) + M^+_B
\]

\setlength{\unitlength}{0.0125in}
\begin{picture}(200,263)(40,540)
\thicklines
\put(300,620){\vector( 0,-1){ 60}}
\put(260,620){\vector( 0,-1){ 60}}
\put(220,620){\vector( 0,-1){ 60}}
\put(180,620){\vector( 0,-1){ 60}}
\put(300,780){\vector( 0,-1){ 60}}
\put(260,780){\vector( 0,-1){ 60}}
\put(220,780){\vector( 0,-1){ 60}}
\put(180,780){\vector( 0,-1){ 60}}
\put(180,780){\framebox(0,0){}}
\put(140,620){\framebox(200,100){}}
\put(300,720){\line(-4,-5){ 80}}
\put(220,720){\line( 4,-5){ 80}}
\put(260,720){\line(-4,-5){ 80}}
\put(180,720){\line( 4,-5){ 80}}
\put (205,550) {\makebox(0,0) [lb] {\raisebox{0pt}[0pt][0pt]{ $M^-_B$}}}
\put (285,785) {\makebox(0,0) [lb] {\raisebox{0pt}[0pt][0pt]{ $M^-_B$}}}
\put (165,550) {\makebox(0,0) [lb] {\raisebox{0pt}[0pt][0pt]{ $M^+_A$}}}
\put (245,785) {\makebox(0,0) [lb] {\raisebox{0pt}[0pt][0pt]{ $M^+_A$}}}
\put (285,550) {\makebox(0,0) [lb] {\raisebox{0pt}[0pt][0pt]{ $M^+_B$}}}
\put (205,785) {\makebox(0,0) [lb] {\raisebox{0pt}[0pt][0pt]{ $M^+_B$}}}
\put (240,550) {\makebox(0,0) [lb] {\raisebox{0pt}[0pt][0pt]{ $M^-_A$}}}
\put (165,785) {\makebox(0,0) [lb] {\raisebox{0pt}[0pt][0pt]{ $M^-_A$}}}
\end{picture}

This ``message switching'' function can be understood in algorithmic terms
as follows.  A demand for output from the application at $M^-_B$ is
switched to the function part of the input, $A \linimpl B$; a demand by the
function input for information about its input at $M_A^-$ is forwarded to
the input port $A$;  a reply with this information about the input at
$M^+_A$ is sent back to the function; an answer from the function to the
original demand for output at $M_B^+$ is sent back to the output port $B$.  
Thus, this strategy does indeed correspond to a protocol for linear
function application---linear in that the ``state'' of the inputs changes as
we interact with them, and there are no other copies available
allowing us to backtrack.  

As for currying, given $\sigma_f: A\tensor B \rightarrow C$, where 
$f: (M^+_A + M^+_B) + M^-_C \rightharpoonup (M^-_A + M^-_B) + M^+_C$,
$\Lambda(\sigma) : A \rightarrow (B \linimpl C)$ is induced by 
\[ M^+_A + (M^+_B + M^-_C) \cong (M^+_A + M^+_B) + M^-_C 
 \stackrel{f}{\rightarrow} (M^-_A + M^-_B) + M^+_C) \cong M^-_A + (M^-_B
+ M^+_C) \]
Finally, note that $A \linimpl {\perp} \cong A^{\perp}$, where this
isomorphism is induced by the bijection 
\[ (M^+_A + \varnothing) + M^-_A  \cong (M^-_A + \varnothing) + M^+_A \]
This yields $(A \linimpl {\perp}) \linimpl {\perp} \cong A^{\perp\perp} = A$.  

\subsubsection{\Games\ as a $\star$-autonomous category}

\begin{proposition}
\Games\ is a $\star$-autonomous category; $\Games_{\tt hf}$ is a
sub-$\star$-autonomous category of \Games.
\end{proposition}
\pr\ We first need to extend the definitions of  $\sigma \tensor \tau$ and
$\Lambda(\sigma)$ from $\Games_{\tt hf}$ to \Games.  This is done as follows.
Let $\sigma: A \rightarrow B$, $\tau: A' \rightarrow B'$.  Then
\[ \sigma \tensor \tau = \{ s \in P_{A \tensor A' \linimpl B \tensor B'}
\mid \; s {\upharpoonright} A,B \in \sigma, \; s {\upharpoonright} A',B'
\in \tau \} \]
We must establish that $\sigma \tensor \tau$ is well-defined and agrees
with the definition in Section~\ref{ghfstar} for history-free strategies.
Firstly, note that, if $ s \cdot c \in \sigma \tensor \tau$
and $c$ is an $O$-move:
\begin{eqnarray*}
c \; \mbox{in $A$ or $B$} \;   \wedge (s \cdot c){\upharpoonright}A,B \cdot d
\in \sigma &\impl& s \cdot c \cdot d \in \sigma \tensor \tau \\
c  \; \mbox{in $A'$ or $B'$} \; \wedge (s \cdot c){\upharpoonright}A',B' \cdot d
\in \tau &\impl& s \cdot c \cdot d \in \sigma \tensor \tau 
\end{eqnarray*}
Now, we show that if $s \in \sigma \tensor \tau$, and $c$ is
an $O$-move in $A$ or $B$ such that $s \cdot c \in P_{A \tensor A' \linimpl
B \tensor B'}$, then the unique $d$ such that $s \cdot c \cdot d \in \sigma
\tensor \tau$ is $\hat{\sigma}((s \cdot c){\upharpoonright}A,B)$; and
similarly if $c$ is in $A'$ or $B'$, with respect to $\tau$.  We argue by
induction on $|s|$; {\em i.e.} we assume the required property for all
proper prefixes of $s$.  It suffices to show that, with the above notation,
if $c$ is in $A$ or $B$, then $d$ in $A'$ or $B'$ and $(s \cdot c \cdot
d){\upharpoonright} A',B' \in \tau$ implies that $d$ is an $O$-move in
$A'$ or $B'$, and hence $s \cdot c \cdot d \not\in \sigma \tensor \tau$.
There are two cases: if $s {\upharpoonright} A',B' = \epsilon$, then $d$
must be an initial move in $\tau$ and hence an $O$-move.  Otherwise,
applying the induction hypothesis to some proper prefix of $s$, the last
$O$-move in $A',B'$ in $s$ must have had its response in $A',B'$ in $s$ and
hence again it is Opponent to move in $s {\upharpoonright} A',B'$ according
to $\tau$.  
 
Let $\sigma: A \tensor B \rightarrow C$.  Then 
$ \Lambda(\sigma) = \{ {\tt assoc}^{\star} (s) \mid s \in \sigma \}$
where 
\[ {\tt assoc}: (M_A + M_B) + M_C \cong M_A + (M_B + M_C) \]
We omit the straightforward verification that this definition agrees with
that of Section~\ref{ghfstar} on history-free strategies.  

At this point, by
Proposition~\ref{hfsubcat} we only
need to show that \Games\ is a $\star$-autonomous category.  We do a sample
calculation below to illustrate the proof.

Firstly, we prove a lemma which halves the work.
\begin{lemma}
Winning strategies are incomparable under inclusion; if $\sigma, \tau$  are
winning strategies in $A$, then $\sigma \subseteq \tau$ implies $ \sigma =
\tau$.  
\end{lemma}
\pr\ Note that any winning strategy $\sigma$ in $A$ satisfies the following
property: if $s \in \sigma$, $O$ to move at $s$, then for all $a$ such that
$ s \cdot a \in P_A$, there is a unique $b$ such that $s \cdot a \cdot b
\in \sigma$. Now, we prove by induction on $|s|$ that $s \in \tau \; \impl\ s
\in \sigma$ .  The base case
$s = \epsilon$ is clear.  Now, suppose $O$ is to move at $s \in \tau$, and
consider any $s \cdot a \in P_A$.  By induction hypothesis, $s \in \sigma$
and since $\sigma, \tau$ are winning, $s \cdot a \cdot b' \in \sigma$ and
$s \cdot a \cdot b'' \in \tau$, for unique $b',b''$.  Since $\sigma
\subseteq \tau$, $s \cdot a \cdot b'' \in \tau$ and $b'=b''$.  Thus, $s
\cdot a \cdot b'' \in \sigma$.    \epr\

Let $\sigma: A \tensor B \rightarrow C$.  We show that the following
diagram commutes (different subscripts have been used on $B,C$ to
distinguish the different occurrences) 
$$
\begin{diagram}
(B_3 \linimpl C_1) \tensor B_2      & \rTo^{{\tt apply}}   & C_2 \\
\uTo^{\Lambda(\sigma) \tensor {\tt id}_B}  &  \NE_{\sigma}                       & \\
A \tensor B_1 & &
\end{diagram}
$$
From the definitions, 
\[
\Lambda(\sigma) \tensor {\tt id}_B; {\tt apply}= \{ s{\upharpoonright}A,B_1,C_2 \mid
s \in S \}
\]
where
\begin{eqnarray*}
S &=& \{ s \in {\cal L}(A \tensor B_1,(B_3 \linimpl C_1)\tensor B_2, C_2)
\mid \\
&&\;\;\;\;\;\;\;\; s {\upharpoonright}A, B_1,B_2,B_3,C_1 \in P_{A \tensor
B_1 \linimpl (B_3 \linimpl C_1)\tensor B_2}, \\
&&\;\;\;\;\;\;\;\; s {\upharpoonright} B_2,B_3,C_1,C_2 \in P_{(B_3 \linimpl
C_1)\tensor B_2 \linimpl C_2} \\
&&\;\;\;\;\;\;\;\; s {\upharpoonright}A,B_3,C_1 \in \sigma \\
&&\;\;\;\;\;\;\;\; s {\upharpoonright}B_1,B_2 \in {\tt id}_B, \; s
{\upharpoonright}B_2,B_3 \in {\tt id}_B  \\
&&\;\;\;\;\;\;\;\; s {\upharpoonright} C_1,C_2 \in {\tt id}_C \}
\end{eqnarray*}

We shall define a map $h$ such that, for all $s \in \sigma$, $h(s) \in S$
and $h(s){\upharpoonright}A,B_1,C_2 =s  $.  This will show that $ \sigma
\subseteq \Lambda(\sigma) \tensor {\tt id}_B; {\tt apply}$, and hence the
desired equation by the above lemma.

We define $h$ as the unique monoid homomorphism extending the following
assignment: 
\begin{description}
\item[$O$-moves: ]
$a  \mapsto a, \;
b \mapsto b_1 \cdot b_2 \cdot b_3, \;
c \mapsto c_2 \cdot c_1$
\item[$P$-moves: ] $
a  \mapsto a, \;
b \mapsto b_3 \cdot b_2 \cdot b_1, \;
c \mapsto c_1 \cdot c_2$
\end{description}
It is clear that for all $s \in \sigma$, $h(s)$ has the following
properties: 
\begin{enumerate}
\item $h(s){\upharpoonright}A,B_1,C_2 = s$
\item $h(s){\upharpoonright}B_1 =h(s){\upharpoonright}B_2  =
h(s){\upharpoonright}B_3$ 
\item $h(s){\upharpoonright}C_1 =h(s){\upharpoonright}C_2$
\item $|s|$ even \impl\ last move in $h(s)$ in $A$, $B_1$ or $C_2$
\end{enumerate}
It remains to show that $h(s) \in S$.  Clearly, (2) applied to all prefixes
of $s$ implies that $h(s){\upharpoonright}B_1, B_2 \in {\tt id}_B$ and
$h(s){\upharpoonright}B_2, B_3 \in {\tt id}_B$.  Similarly, (3) implies that
$h(s){\upharpoonright}C_1, C_2 \in {\tt id}_C$.  Also, (1), (2) and (3) and $s
\in \sigma$ implies that $h(s){\upharpoonright}A,B_3,C_1 \in \sigma$. 

Now, let $t = h(s) {\upharpoonright}A, B_1,B_2,B_3,C_1$, $ T = P_{A \tensor
B_1 \linimpl (B_3 \linimpl C_1)\tensor B_2}$.  We will show that $t \in T$,
by induction on 
$|s|$.  For the key case, suppose Opponent to move at $s$.  Let $s \cdot d
\cdot e \in \sigma$.  We now consider the various subcases according to the
locations of $d$ and $e$.  For example, suppose $d=b$ is in $B$, and $e =
c$ is in $C$.  Then 
$ h(s \cdot b \cdot c) = h(s) \cdot b_1 \cdot b_2 \cdot b_3 \cdot c_1
\cdot c_2 $ and
$  h(s \cdot b \cdot c){\upharpoonright} A, B_1,B_2,B_3,C_1 =
t \cdot b_1 \cdot b_2 \cdot b_3
\cdot c_1$.  
By induction hypothesis, $t \in T$.  By (1), (4) and
$s \cdot d \in \sigma$, 
$ t \cdot b_1 \in T$.  Using (2), $
t \cdot b_1 \cdot b_2 \cdot b_3\in T $.  
Using (1), (3) and $s \cdot d \cdot e \in \sigma$, we get the required
result.  A similar argument shows that $
h(s){\upharpoonright}B_2,B_3,C_1,C_2 \in P_{(B_3 \linimpl C_1)\tensor B_2
\linimpl C_2} $.  Also, note that if $s \cdot d \cdot e \in \sigma$,
where $d$ is in $A$ or $B$ and $e$ is in $C$, then $e$ must be a $P$-move;
similarly, if $d$ is in $C$ and $e$ is in $A$ or $B$.  It then easily
follows, by induction on $|s|$, that $ h(s) \in {\cal L}(A \tensor B_1,(B_3 \linimpl C_1)\tensor
B_2, C_2) $.

We also verify the unicity equation $\Lambda(\tau \tensor {\tt
id}_B;{\tt apply}) = \tau$, where $\tau: A \rightarrow (B \linimpl C)$.  We
define  $ \Lambda^{-1}(\tau)  = \{ ({\tt assoc}^{-1})^{\star}(s) \mid s \in
\tau \}$.   
Clearly, $\Lambda (\Lambda^{-1} (\tau)) = \tau$ and $\Lambda^{-1} (\tau): A
\tensor B \rightarrow C$.  Now, 
\begin{eqnarray*}
\Lambda(\tau \tensor {\tt id}_B;{\tt apply})&=& \Lambda(\Lambda (\Lambda^{-1}
(\tau)) \tensor {\tt id}_B;{\tt apply}) \\
&=& \Lambda (\Lambda^{-1} (\tau)) \\
&=& \tau. 
\end{eqnarray*}
\epr

\subsection{Variable types and uniform strategies}\label{vartypes}
An {\em embedding } $e: A \rightarrowtail B$ is a 1--1 map $e: M_A \rightarrow
M_B$ such that
\begin{description}
\item[(e1)] $\lambda_B \circ e = \lambda_A$
\item[(e2)] $e^{\star}(P_A) \subseteq P_B$
\item[(e3)] $(\forall s \in P^{\infty}_A) \; [ s \in W_A \; \ifof\
e^{\omega}(s) \in W_B]$
\end{description}
where $e^{\star},e^{\omega}$ are the canonical extensions of $e$ to
$M^{\star}_A,M^{\omega}_A$ respectively.  We write $\Games^e$ for the
evident category of games and embeddings.  Note that given an embedding $e:
A \rightarrowtail B$, we can derive functions $e^-:
M_A^- \rightarrowtail M^-_B$ and $e^+:
M_A^+ \rightarrowtail M^+_B$. 

\begin{proposition}
Tensor, Par and Involution can be extended to covariant functors over
$\Games^e$.
\end{proposition}
\pr\ If $e: A \rightarrowtail B, \; e': A' \rightarrowtail B'$, then
$e\tensor e' = e+ e'$ and $e^{\perp} =e$.  We just check the only
non-obvious part, namely that condition (e3) is satisfied by $e^{\perp}$.
Given $s \in P^{\infty}_{A^{\perp}} = P^{\infty}_{A}$,
\begin{eqnarray*}
 s \in W_{A^{\perp}} &\ifof& s \in   P^{\infty}_{A} \setminus W_A \\
&\ifof& s \in   P^{\infty}_A, \; e^{\omega}(s) \not\in W_B \\
&\ifof& s \in   P^{\infty}_{A^{\perp}}, \; e^{\omega}(s) \in W_{B^{\perp}}
\end{eqnarray*}
Thus, $ s \in W_{A^{\perp}} \; \ifof\ e^{\omega}(s) \in W_{B^{\perp}}$. \epr 

Now, given a multiplicative formula $A$ with propositional atoms $\alpha_1,
\ldots, \alpha_n$, this induces a functor $F_A: (\Games^e)^n \rightarrow
\Games^e$.  Similarly, a sequent $\Gamma(\alpha_1,\ldots,\alpha_n)$ induces
a functor $F_{\Gamma}: (\Games^e)^n \rightarrow \Games^e$ (where
$\Gamma$ is interpreted as $\parc \Gamma$).

A strategy for $\Gamma(\alpha_1,\ldots,\alpha_n)$ will be a family $\{
\sigma_{\vec{A}} \}$, where for each n-tuple of games $\vec{A} $,
$\sigma_{\vec{A}} $ is a strategy in $F_{\Gamma}(\vec{A})$.  We express
the uniformity of this family by a naturality condition.  Given $F:
(\Games^e)^n \rightarrow \Games^e$, we define two functors $F^-, F^+:
(\Games^e)^n  \rightarrow {\tt Set}^p$, where ${\tt Set}^p$ is the category
of sets and partial functions.  
\[ \begin{array}{ccc}
F^-(\vec{A}) = M^-_{F(\vec{A})} &\hspace{1.5in}& F^-(\vec{e}) = F(\vec{e})^-  \\
F^+(\vec{A}) = M^+_{F(\vec{A})}&& F^+(\vec{e}) = F(\vec{e})^+
\end{array}
\]
If $\sigma = \{ \sigma_{\vec{A}} \}$ is a family of history free
strategies, then each $\sigma_{\vec{A}}$ is of the form
$\sigma_{f_{\vec{A}}}$.  So we get a family of partial functions $\{
f_{\vec{A}} \}$ where
$ f_{\vec{A}} :  M^-_{F(\vec{A})} \rightarrow M^+_{F(\vec{A})}$, {\em i.e.}
$f_{\vec{A}}:  F^-(\vec{A}) \rightarrow F^+(\vec{A})$.  
We say that $\sigma$ is {\em uniform} if $f$ is a natural transformation
$f: F^- \rightarrow F^+$.

Now, for each $n \in \omega$, we can define a category $\Games_{\tt hf}(n)$,
whose objects are functors $F: (\Games^e)^n \rightarrow \Games^e$ and whose
morphisms $\sigma: F \rightarrow G$ are uniform, history-free winning strategies
$\{ \sigma_{\vec{A}} \}$, where $\sigma_{\vec{A}}: F(\vec{A})  \rightarrow
G(\vec{A})$, {\em i.e.} $\sigma_{\vec{A}}$ is a strategy in $F(\vec{A})\;
\linimpl\ G(\vec{A})$.  Composition is pointwise: if $\sigma: F \rightarrow
G, \; \tau: G \rightarrow H$, then $(\sigma; \tau)_{\vec{A}} =
\sigma_{\vec{A}};\tau_{\vec{A}}$.  Note that $\Games_{\tt hf}(0) \cong
\Games_{\tt hf}$.  

\begin{proposition}
For each $n$, $\Games_{\tt hf}(n)$ is a $\star$-autonomous category;
$\Games_{\tt hf}: {\Bbb B}
\rightarrow \star$-{\bf Aut} is an indexed $\star$-autonomous category with
base ${\Bbb B}$, the category of finite ordinals and set maps.
\end{proposition}
\pr\ The $\star$-autonomous structure on $\Games_{\tt hf}(n)$ is defined
pointwise from that on $\Games_{\tt hf}$, {\em e.g.} $(F \tensor
G)(\vec{A}) = F(\vec{A}) \tensor G(\vec{A})$. 

We will show that composition preserves uniformity.  Given functions $f,g$
as in Section~\ref{gamesgoi}, we write ${\tt EX}(f,g)$ for the execution
formula applied to $f,g$.  Now, if $\sigma: F \rightarrow G$, $\tau: G
\rightarrow H$, $\sigma = \sigma_f$ and $\tau = \tau_g$, and $\vec{e}:
\vec{A} \rightarrowtail \vec{B}$, we must show that

$$
\begin{diagram}
M^-_{F(\vec{A}) \linimpl H(\vec{A})} && \rTo^{{\tt
EX}(f_{\vec{A}},g_{\vec{A}})}   && M^+_{F(\vec{A}) \linimpl H(\vec{A})} \\ 
\dTo^{(F \linimpl H) (\vec{e})^-}  &&  &&    \dTo_{(F \linimpl H) (\vec{e})^+}\\
M^-_{F(\vec{B}) \linimpl H(\vec{B})} && \rTo_{{\tt
EX}(f_{\vec{B}},g_{\vec{B}})}   && M^+_{F(\vec{B}) \linimpl H(\vec{B})} \\
\end{diagram}
$$
Writing ${\displaystyle {\tt
EX}(f_{\vec{A}},g_{\vec{A}}) = \bigvee_{k \in \omega} m_k^{\vec{A}}}$
where 
${\displaystyle  m_k^{\vec{A}} = \pi^{\star}_{\vec{A}} \circ ((f_{\vec{A}}
+ g_{\vec{A}}) \circ \mu_{\vec{A}})^k \circ (f_{\vec{A}} + g_{\vec{A}}) \circ
\pi_{\vec{A}} }$,
we must show that
\[ (F(\vec{e})^- + H(\vec{e})^+) \circ \bigvee_{k \in \omega} m^{\vec{A}}_k
= \bigvee_{k \in \omega} m_k ^{\vec{B}}\circ (F(\vec{e})^+ + H(\vec{e})^-) \]
Since composition distributes over joins, it suffices to show that for all
$k$, 
\begin{eqnarray}
(F(\vec{e})^- + H(\vec{e})^+) \circ m^{\vec{A}}_k &=& m_k^{\vec{B}} \circ
(F(\vec{e})^+ + H(\vec{e})^-) \label{indn}
\end{eqnarray}
Note firstly that
\begin{eqnarray*}
(F(\vec{e})^- + H(\vec{e})^+) \circ \pi^{\star}_{\vec{A}} &=&
\pi^{\star}_{\vec{B}} \circ (F(\vec{e})^- + G(\vec{e})^+ + G(\vec{e})^- +
H(\vec{e})^+ ) \\
\pi_{\vec{B}} \circ (F(\vec{e})^+ + H(\vec{e})^-) &=& (F(\vec{e})^+ +
G(\vec{e})^- + G(\vec{e})^+ + H(\vec{e})^- ) \circ \pi_{\vec{A}} \\
(F(\vec{e})^+ + G(\vec{e})^- + G(\vec{e})^+ + H(\vec{e})^- ) \circ
\mu_{\vec{A}} &=& \mu_{\vec{B}} \circ (F(\vec{e})^- + G(\vec{e})^+ +
G(\vec{e})^- + H(\vec{e})^+ ) 
\end{eqnarray*}
and by uniformity of $f$ and $g$
\[ (f_{\vec{B}} + g_{\vec{B}} ) \circ (F(\vec{e})^+ + G(\vec{e})^- +
G(\vec{e})^+ + H(\vec{e})^-) = (F(\vec{e})^- +
G(\vec{e})^+ + G(\vec{e})^- +
H(\vec{e})^+) \circ (f_{\vec{A}} + g_{\vec{A}} ) \]
A straightforward induction on $k$ using these equations
establishes (\ref{indn}). 

The uniformity of the morphisms witnessing the $\star$-autonomous structure
on $\Games_{\tt hf}(n)$ follows directly from the naturality of the
canonical isomorphisms for coproduct in ${\tt Set}$ from which they are
defined.  

Given $f: \{ 1, \ldots,n \} \rightarrow \{ 1,\ldots,m \}$ (where we take
the liberty of representing the ordinal $n$ by $\{ 1, \ldots,n \}$), we define 
\begin{eqnarray*}
 \Games_{\tt hf}(f)(F)(A_1,\ldots,A_m) &=& F(A_{f(1)},\ldots,A_{f(n)}) \\
\Games_{\tt hf}(f)\{ \sigma_{A_1,\ldots,A_n} \} &=& \{ \sigma_{ A_{f(1)},
\ldots, A_{f(n)}} \}
\end{eqnarray*}
The verification that $\Games_{\tt hf}(f)$ is a $\star$-autonomous functor
is straightforward from the pointwise definition of the $\star$-autonomous
structure on $\Games_{\tt hf}(n)$.  The functoriality of $\Games_{\tt hf}$
itself is a routine calculation.   \epr\  

Using this Proposition, we can interpret proofs in MLL + MIX by uniform,
history-free strategies; see~\cite{Seeley} for further details.  This is
the semantics for which Full Completeness will be proved.

\section{Full Completeness}\label{proofs}

In this section, we prove full Completeness of the game semantics for MLL
+ MIX.  The proof is structured into a number of steps.

\begin{itemize}
\item Firstly, we show that a uniform, history free winning strategy
for $\Gamma$ induces a proof structure on $\Gamma$.

\item Next, we reduce the problem to that for {\em binary} sequents,
in which each atom occurring does so once positively
and once negatively.

\item We then make a further reduction to {\em simple} binary sequents, in
which every formula is either a literal, or the tensor product of two
literals.

\item Finally, we show that for such sequents, there can only be a winning
strategy if the corresponding proof structure satisfies the correctness
criterion, {\em i.e.} is a proof net. 

\end{itemize}

\subsection{Strategies induce Axiom links}
We begin by establishing some notation.  We are  given an MLL sequent
$\Gamma(\alpha_1,\ldots,\alpha_k)$ where $\alpha_1,\ldots,\alpha_k$ are the
propositional atoms occurring in $\Gamma$.  We enumerate the occurrences of
literals in $\Gamma$ as $c_1,\ldots c_n$; each $c_i$ is an occurrence of
$l_i$, where $l_i = \alpha_{j_i}$ or $l_i = \alpha_{j_i}^{\perp}$ for some
$j_i$, $1 \leq i \leq n, 1 \leq j_i \leq k$.  Given any sequence $\vec{A} =
A_1,\ldots,A_k$ of games instantiating $\alpha_1, \ldots,\alpha_k$, we
obtain a game $F(\vec{A})$, where $F = F_{\Gamma}$ is the interpretation of
$\parc \Gamma$.  Note that $M_{F(\vec{A})} = \sum^n_{i=1} M_{C_i} $,
where $C_i = A_{j_i}$ or $A_{j_i}^{\perp}$.  We represent
$M_{F(\vec{A})}$ concretely as $\cup^n_{i=1} \{i\} \times M_{C_i}$.
We refer to the $C_i$ as the {\em constituents} of $M_{F(\vec{A})}$.

\begin{proposition}\label{prop1}
With notation as above, let $\sigma = \{ \sigma_A \}$ be a uniform history
free winning strategy for $F = F_{\Gamma}$.  Then, for some
involution $\phi$ such that $(\Gamma, \phi)$ is a proof structure, for all $\vec{A}$,
\[ \sigma_{\vec{A}} = \sigma_{f_{\vec{A}}} \]
where $f_{\vec{A}}((i,a)) = (\phi(i),a )$.
\end{proposition}
\pr\ A game $A$ is {\em full} if $P_A = M_A^{\circledast}$.  Given any
game $A$, there is an embedding $e^{\tt full}_A: A \rightarrowtail A^{\tt full}$, where $A^{\tt full}
= (M_A, \lambda_A, M_A^{\circledast},W_A)$ and $e^{\tt full}_A = {\tt id}_{M_A}$.  

By uniformity,

$$
\begin{diagram}
F^{-}(\vec{A})      && \rTo^{F^{-}(e^{\tt full}_{\vec{A}})}   && F^{-}(\vec{A}^{\tt full}) \\
\dTo^{f_{\vec{A}}}  &&                               && \dTo_{f_{\vec{A}^{\tt full}}} \\
F^{+}(\vec{A})      && \rTo_{F^{+}(e^{\tt full}_{\vec{A}})}      && F^{+}(\vec{A}^{\tt full})
\end{diagram}
$$

But $\negg{F}(e^{\tt full}_A) = {\tt id}_{M^{-}_{F(A)}}, \; \posg{F}(e^{\tt full}_A) =
{\tt id}_{M^{+}_{F(A)}}$.  Hence $f_{\vec{A}} = f_{\vec{A}^{\tt full}}$.  Thus, it
suffices to prove the Proposition for full games.

Let $i \in \{1,\ldots,n
\}$ and $a \in M^{-}_{C_i}$.  Thus, $(i,a)$ is an $O$-move in the $i$'th
constituent of $F(\vec{A})$.  Consider the vector $\vec{B}$, where the
$i$'th constituent is instantiated with
\[ B  = ( \{ b \}, \{(b,O)\}, \{ \epsilon, b \}, \varnothing ), \]
all constituents labelled with the same literal by $B$, all constituents
labelled with the dual literal by $B^{\perp}$, all other constituents with
the empty game.  Since $\sigma_{\vec{B}}$ is winning, we must have
$f_{\vec{B}} ((i,b)) = (j,b)$, for some constituent $j$ with dual label to
that of $i$.

Now there is an embedding from $B$ to $A_{j_i}$, hence from $\vec{B}$ to
$\vec{A}$, sending $b$ to $a$.  By uniformity, this implies that
$f_{\vec{A}}((i,a)) = (j,a)$.  Note that this will apply to {\em all}
$(i,a')$ for the given $i$, so all $O$-moves in the $i$'th constituent are
mapped to the {\em same} fixed constituent $j$.  Thus, we can define an
endofunction $\phi$ on $\{ 1, \ldots, n \}$ such that, for all full $\vec{A}$, and
hence for all $\vec{A}$, for all $i \in \{ 1, \ldots, n \}, \; a \in
M_{A_{j_i}}^{-}, f_{\vec{A}}((i,a)) = (\phi(i),a)$.  Moreover, $l_i =
l_{\phi(i)}^{\perp}$, so in particular $\phi$ is fixpoint free.

It only remains to be shown that $\phi$ is an involution. Consider the game

\[ C = ( \{ a',b' \}, \{(a',O),(b',P)\}, \{ \epsilon, a', a'\cdot b' \},
\varnothing) \]
Consider the instance $\vec{C}$ defined similarly to $\vec{B}$, with $C$
used in place of $B$.  We already know that $f_{\vec{C}}((i,a')) =
(\phi(i),a')$.  Since $\sigma_{\vec{C}}$ is winning, we must have
$f_{\vec{C}}((\phi(i),b')) = (i,b')$.  So $\phi^2(i) =i$, and $\phi$ is an
involution as required.  \epr\

\begin{cor}
If there is a uniform history-free winning strategy for $F = F_{\Gamma}$,
then $\Gamma$ must be balanced, {\em i.e.} each atom must occur the same
number of times positively as negatively.  
\end{cor}
\pr\  The function $\phi$ of Proposition~\ref{prop1} establishes a
bijection between positive and negative occurrences of each atom. 
\epr\

\subsection{Reduction to binary sequents}

Let $\sigma$ be a history free strategy for a proof structure $(\Gamma,
\phi)$.  We define a binary sequent $\Gamma_{\phi}$ by relabelling the
literals using distinct atoms except that each $i$ remains dual to
$\phi(i)$.  Note that a binary sequent has a unique associated proof
structure; so the involution is redundant in this case.  It is clear from
the definition of the correctness criterion that
\[ (\Gamma, \phi) \; \mbox{ is a proof net} \; \ifof\  \Gamma_{\phi} \mbox{ is
a proof net} \]
Now given a proof structure $(\Gamma, \phi)$, the corresponding uniform,
history-free strategy $\sigma_{(\Gamma, \phi)}$ for $\Gamma$ is defined by 
\[ \sigma_{(\Gamma, \phi)} = \sigma_{f_{(\Gamma, \phi)}}, \; \mbox{where} \;
f_{(\Gamma, \phi),\vec{A}}((i,a))  = (\phi(i),a) \]

\begin{proposition}\label{tobinary}
Let $(\Gamma, \phi)$ be a proof structure.  
\[ \sigma_{(\Gamma, \phi)} \; \mbox{is winning for} \; \Gamma
\ifof\ \sigma_{\Gamma_{\phi}} \; \mbox{ is winning for} \; \Gamma_{\phi} \]
\end{proposition}
\pr\ Since every instance of $\Gamma$ is an instance of $\Gamma_{\phi}$, the
right to left implication is clear.

For the converse, given an instance $\vec{A}$ for $\Gamma_{\phi}$, consider
the following instance for $\Gamma$: for each $\alpha$ occurring $k$ times
positively in $\Gamma$, with $A_{j_1}, \ldots, A_{j_k}$ instantiating these
occurrences in $\vec{A}$, instantiate $\alpha$ with the disjoint union $
A_{j_1}+\cdots+A_{j_k}$.  Since $\sigma_{(\Gamma, \phi)}$ is winning by
assumption, it defeats every play by Opponent, in particular those plays in
which Opponent plays only in $A_{j_i}$ in the game instantiating the $i$'th
occurrence of $\alpha$.  This shows that $\sigma_{\Gamma_{\phi}}$ is
winning as required.  

\epr\

\subsection{Reduction to simple sequents}
Let $\Gamma$ be a binary sequent.  We write $\Gamma = D[A]$, where
$D[\cdot]$ is a monotone context, {\em i.e.} with the ``hole'' $[\cdot]$
appearing only under the scope of Tensors and Pars.  For such a context, we
have
\[ A \linimpl B \entails D[A] \linimpl  D[B] \]

\begin{lemma}\label{prop21}
Let $\Gamma = D[A \tensor (  B \parc C)]$ be a binary sequent.  Let 
$\Gamma_1 = D [ (A \tensor B) \parc C]$ and   
$\Gamma_2 = D [ (A \tensor C) \parc B] $ .  Then 
\begin{enumerate}
\item $(\forall i) \;  \entails \Gamma \linimpl \Gamma_i $
\item $\entails \Gamma  \; \ifof\   (\forall i) \; \entails \Gamma_i$
\end{enumerate}
\end{lemma}
\pr\  \hfill
\begin{enumerate}
\item $A \tensor (  B \parc C) \linimpl (A \tensor B) \parc C$
and $A \tensor ( B \parc C) \linimpl (A \tensor C) \parc B$ are both
theorems of MLL.  
\item We use the correctness
criterion.   Suppose $\Gamma$ is not provable, {\em i.e.} for some
switching $S$, $G(D[A \tensor (  B \parc C)], S)$ has a
cycle.  If $S$ sets the indicated par link to $L$, there will be a cycle in
$\Gamma_1$; if $S$ sets the indicated par link to $R$, there will be a cycle in
$\Gamma_2$.  \epr\
\end{enumerate}

\begin{lemma}\label{prop22}
Let $\Gamma = D[A \tensor (  B \tensor C)]$ be a binary sequent.  Let 
$\Gamma_1 = D [ A \tensor (B \parc C)],\; \Gamma_2 = D [ A \parc (B \tensor
C)]$.  Then, 
\begin{enumerate}
\item $(\forall i) \; \entails \Gamma \linimpl \Gamma_i $
\item $\entails \Gamma  \; \ifof\   (\forall i) \; \entails \Gamma_i$
\end{enumerate}
\end{lemma}

\pr\  \hfill
\begin{enumerate}
\item  $\alpha \tensor \beta  \linimpl \alpha \parc \beta$ is a theorem of
MLL + MIX.

\item We use the correctness
criterion.  Suppose $\Gamma$ is not provable, {\em i.e.} for some switching $S$
$G(\Gamma, S)$ has a
cycle.  In particular fix some {\em simple} cycle in $G(\Gamma, S)$ ({\em
i.e.} no internal node is visited more than once).  This implies that the
cycle cannot visit all of the $A,\;B,\;C$ edges.  Thus, there are four
possible cases:
\begin{itemize}
\item  The cycle does not visit $A \tensor (  B \tensor C)$ at all. Then
clearly both $\Gamma_1,\; \Gamma_2$ have cycles.

\item The cycle visits the $A$ and $B$ edges:  Then $G(\Gamma_1, S')$ has a
cycle, where $S'$ sets the switch of the new Par node to $L$, and
otherwise is defined like $S$.

\item The cycle visits the $A$ and $C$ edges:  Symmetric to the previous case.

\item The cycle visits the $B$ and $C$ edges:  Then $G(\Gamma_2, S')$ has
a cycle, where $S'$ sets the switch of the new Par node to $R$, and 
otherwise is defined like $S$.   \epr\
\end{itemize}
\end{enumerate}

\begin{proposition}\label{simplifybinary}
Let $\Gamma$ be a binary sequent.  Then there is a set of {\em simple}
binary sequents $\Gamma_1, \ldots, \Gamma_n$ such that:
\begin{enumerate}
\item $(\forall i) \; \entails \Gamma \linimpl \Gamma_i$
\item $\entails \Gamma_{\phi} \; \ifof\ (\forall i) \;  \entails \Gamma_i $
\end{enumerate}

\end{proposition}
\pr\ Firstly, use Lemma~\ref{prop21} repeatedly to push all Pars to
the top and then replace them by commas.  Then, given a nested occurrence
of Tensor, we can use Lemma~\ref{prop22} to replace it with a Par, and use
Lemma~\ref{prop21} again to eliminate this Par.  In this way, we eventually
reach a set of {\em simple} binary sequents.  \epr\

\subsection{Winning strategies are acyclic}

We now establish the crucial connection between winning strategies and the
correctness criterion for proof nets. 

\begin{proposition}\label{invalidbinary}
Let $\Gamma$ be a {\em simple} binary sequent.  Let $\sigma_{\Gamma}$ be
the associated uniform history free strategy as in
Proposition~\ref{tobinary}.  If $\sigma_{\Gamma}$ is winning, then the
(unique) proof structure associated with
$\Gamma$ is acyclic. 
\end{proposition}
\pr\  Suppose $\Gamma$ has a cycle.  Since $\Gamma$ is simple, this is
necessarily of the form

\[l_1^{\perp},\tensor,l_2,l_2^{\perp}, \tensor,\ldots,
l_n,l_n^{\perp},\tensor, l_1 \]
For example:  
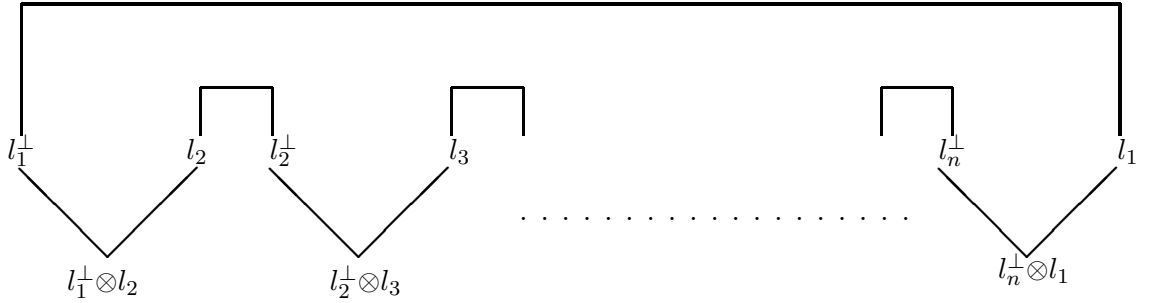
\begin{center}
\setlength{\unitlength}{0.0125in}
\begin{picture}(541,135)(100,690)
\thicklines
\put(500,765){\line( 0, 1){ 20}}
\put(500,785){\line(-1, 0){ 30}}
\put(470,785){\line( 0,-1){ 20}}
\put(320,765){\line( 0, 1){ 20}}
\put(320,785){\line(-1, 0){ 30}}
\put(290,785){\line( 0,-1){ 20}}
\multiput(320,730)(8.88889,0.00000){19}{\makebox(0.4444,0.6667){.}}
\put(110,765){\line( 0, 1){ 55}}
\put(110,820){\line( 1, 0){460}}
\put(570,820){\line( 0,-1){ 55}}
\put(215,765){\line( 0, 1){ 20}}
\put(215,785){\line(-1, 0){ 30}}
\put(185,785){\line( 0,-1){ 20}}
\put(146,714){\line( 1, 1){ 37.500}}
\put(146,714){\line(-1, 1){ 37}}
\put(251,714){\line(-1, 1){ 37}}
\put(251,714){\line( 1, 1){ 37.500}}
\put(531,714){\line( 1, 1){ 37.500}}
\put(531,714){\line(-1, 1){ 37}}
\put (235,700) {\makebox(0,0) [lb] {\raisebox{0pt}[0pt][0pt]{ $l^{\perp}_2 \tensor l_3$}}}
\put (125,700) {\makebox(0,0) [lb] {\raisebox{0pt}[0pt][0pt]{ $l_1^{\perp} \tensor l_2$}}}
\put (515,705) {\makebox(0,0) [lb] {\raisebox{0pt}[0pt][0pt]{ $l_n^{\perp} \tensor l_1$}}}
\put (565,755) {\makebox(0,0) [lb] {\raisebox{0pt}[0pt][0pt]{ $l_1$}}}
\put (490,755) {\makebox(0,0) [lb] {\raisebox{0pt}[0pt][0pt]{ $l_n^{\perp}$}}}
\put (285,755) {\makebox(0,0) [lb] {\raisebox{0pt}[0pt][0pt]{ $l_3$}}}
\put (210,755) {\makebox(0,0) [lb] {\raisebox{0pt}[0pt][0pt]{ $l^{\perp}_2$}}}
\put (175,755) {\makebox(0,0) [lb] {\raisebox{0pt}[0pt][0pt]{ $l_2$}}}
\put (100,755) {\makebox(0,0) [lb] {\raisebox{0pt}[0pt][0pt]{ $l_1^{\perp}$}}}
\end{picture}

\end{center}
(This picture is not completely general; non-planar arrangements are
possible.  However, this will not play any role
in the argument).

We will assign games $\vec{A}$ to atoms in $\Gamma$ in such a way that
Opponent has a winning strategy in $F_{\Gamma}(\vec{A})$, thus showing that
there can be no uniform winning strategy for $\Gamma$.

We label the literals $l_1^{\perp},l_2,l_2^{\perp},\ldots,
l_n,l_n^{\perp}, l_1$ alternately {\em tt} and {\em ff}.  We define
$\vec{A}$ such that each literal labelled {\em tt} is assigned
\[ (\{ a \}, \{ (a, P) \},\{  a , \epsilon \}, \varnothing) \]
and each literal labelled {\em ff} is assigned 
\[ (\{ a \}, \{ (a, O) \},\{  a , \epsilon \}, \varnothing) \]
and all unlabelled literals are assigned the empty game. 

We now describe the strategy for Opponent.  Note that by assumption, Player
is following the strategy $\sigma_{\Gamma}$, so his response to Opponent's
moves is determined a priori.

Consider the following play:
\begin{center}
\begin{tabular}{l}
$O$ plays $a$  in $l_1$ \\
$P$ plays $a$  in $l_1^{\perp}$ \\
$O$ plays $a$  in $l_2$ \\
$P$ plays $a$  in $l_2^{\perp}$\\
$ \vdots $ \\
$O$ plays $a$  in $l_n$ 
\end{tabular}
\end{center}
By strategy $\sigma_{\Gamma}$, $P$ has to play plays $a$  in $l_n^{\perp}$.
Note that the only previous move in the subgame $l_n^{\perp} \tensor l_1$
was $O$'s opening move in $l_1$. Thus, $P$'s move would switch to the
other side of the tensor, which is prohibited by the rules governing the
valid positions for tensor. Hence, $P$ loses this play.  \epr\

\subsection{Main result}

\begin{thm} (Full Completeness) \\
If $\sigma$ is a uniform history-free winning strategy for $\Gamma$, then
it is the denotation of a unique proof net $(\Gamma, \phi)$.
\end{thm}
\pr\ By Proposition~\ref{prop1}, we know that there is a unique proof
structure $(\Gamma, \phi)$ with $\sigma = \sigma_{(\Gamma,\phi)}$.  It
remains to show that  $(\Gamma, \phi)$ is a proof net.  By
Proposition~\ref{tobinary}, $\sigma_{(\Gamma,\phi)}$ winning implies
$\sigma_{\Gamma_{\phi}}$ winning.  Applying
Proposition~\ref{simplifybinary} to $\Gamma_{\phi}$, there is a set of
simple binary sequents $\Gamma_1, \ldots, \Gamma_n$ such that
\begin{enumerate}
\item $(\forall i) \;  \entails \Gamma_{\phi} \linimpl \Gamma_i $
\item $\entails \Gamma_{\phi} \; \ifof\ (\forall i) \;  \entails \Gamma_i $
\end{enumerate}
Since the game semantics is sound, (1) and the validity of $\Gamma_{\phi}$
in the game semantics implies that there is a uniform, history-free winning
strategy for each $\Gamma_i$.  By Proposition~\ref{prop1}, this strategy is
necessarily of the form $\sigma_{\Gamma_i}$.  By
Proposition~\ref{invalidbinary}, this implies that each $\Gamma_i$ is
acyclic.  By (2), this implies that $\Gamma_{\phi}$ is a proof net.  By the
remark before Proposition~\ref{tobinary}, this implies that $(\Gamma,
\phi)$ is a proof net.  \epr\

\section{Beyond the multiplicatives}\label{Fullinterpretation}

Up to this point, we have only considered the multiplicative fragment of
Linear Logic.  However, our game semantics in fact yields a categorical
model of full second-order (or even $\omega$-order) Classical Linear
Logic.  In this section, we will outline the interpretation of the
additives and exponentials.  A detailed treatment of this material, and of
the game semantics for the second-order quantifiers, will be given in a
sequel to the present paper.

\subsection{Polarities}
To proceed, we focus on the fact that our games may admit some positions in
which Player starts, some in which Opponent starts.  

\begin{df}
A game $A$ is {\em positive} (has polarity +1) if every valid initial move
in $A$ is by Player; {\em negative} (has polarity $-1$) if every valid initial
move in $A$ is by Opponent; and {\em neutral} (polarity $0$) otherwise.
\end{df}

Although we use the same notation for polarities as
Girard~\cite{GirU90}, they have a somewhat different interpretation.
Our polarities have a very direct computational reading.  If we interpret
moves by Opponent as demands for data and moves by Player as
generating data, then {\em positive} games model purely data-driven
computation; {\em negative} games model purely demand-driven computation;
while {\em neutral } games allow both modes of computation.  These notions
give rise to the following situation.
We have full subcategories 
\[ \negg{I}: \negg{\Games} \hookrightarrow \Games \hookleftarrow
\posg{\Games}: \posg{I} \]
of positive and negative games.  There are evident constructions \posg{A}
(\negg{A}) taking a game $A$ in \Games\ to \posg{\Games} (\negg{\Games})
simply by deleting all positions of $P_A$ starting with a move by Opponent
(Player) and correspondingly pruning $W_A$.  

\begin{proposition}\label{proppol} \hfill 
\begin{itemize}
\item \posg{\Games} is reflective and \negg{\Games} is co-reflective in \Games,
with $\negg{I} \dashv \negg{(\cdot)}, \; \posg{(\cdot)} \dashv \posg{I}$.
\item Linear negation $(\cdot)^{\perp}$ cuts down to a duality
$\negg{\Games} \simeq \posg{\Games}^{\tt op}$; in fact $(\negg{A})^{\perp}
=\posg{(A^{\perp})}, \; (\posg{A})^{\perp}
=\negg{(A^{\perp})}$. 
\end{itemize}
\end{proposition}

\subsection{Exponentials}

Jacobs has recently investigated the decomposition of the exponentials
$\ofcourse, \; \whynot$ into weakening parts $\ofcourse_w, \; \whynot_w$
and contraction parts $\ofcourse_c, \; \whynot_c$~\cite{Jac92}.  He
develops a general theory for this decomposition.  We will use a little of
this theory to structure our presentation of the exponentials.

\subsubsection{Weakening}
The reflection and co-reflection of Proposition~\ref{proppol} give rise
to a monad and a comonad on \Games\ respectively, which we denote by
$\whynot_w$ and $\ofcourse_w$.  Our reason for this notation is explained
by the following proposition.

\begin{proposition}
There are natural transformations 
\[ \ofcourse_w A \tensor B \rightarrow B, \;\; B \rightarrow \whynot_w A
\parc B \]
\end{proposition}
As a consequence of this proposition, the following weakening rule is valid
in the game semantics.

\[ \mytwotrans{ \entails \Gamma} { \entails \Gamma, \whynot_w A } \]

\subsubsection{Exponentials}
We want to define $\ofcourse A$ as the type of objects which are copyable
versions of objects of type $A$.  We achieve copyability by {\em
backtracking}; {\em cf.}~\cite{AV91}.  That is, at any stage in a play in
$\ofcourse A$, the Opponent may return to a previous stage to make his
move.  In this way, a single play in $\ofcourse A$ will correspond to a
tree of plays in $A$.

\begin{df}
$\ofcourse A$ is defined as follows:
\begin{itemize}
\item $M_{\ofcourse A} = M^+_A \cup\ (\omega \times M^-_A)$
\item $\lambda_{\ofcourse A}(a) = P, \; \; \; \; \lambda_{\ofcourse
A}((i,a)) = O$
\item Define 
\begin{itemize}
\item $s(i) = s_1 \cdots s_i, \; s {\setminus} i = s_1 \cdots s_{|s| -
i}$
\item $ \overline{(\;)}: M_{\ofcourse A}^{\star} \rightarrow
M_{\ofcourse A}^{\star}$ by 
$ \overline{\epsilon} = \epsilon, \; \overline{s \cdot a} =
\overline{s}\cdot a, \; \overline{s \cdot (i,a)} = \overline{(s {\setminus} i)}
\cdot a $
\item $\hat{s} = \{ \overline{s(i)} \mid 0 \leqslant i \leqslant |s| \}$
\end{itemize}
Also, a {\em partial strategy} is defined like a strategy except that it need
not satisfy {\bf (s3)}.
Then, 
\[ P_{\ofcourse A} = \{ s \in M^{\circledast}_{\ofcourse A} \mid
(\forall j: 1 \leqslant j \leqslant |s|) \; s_j = (i,a) \; \impl\ i < j, \;\hat{s} \mbox{
is a partial strategy in} \; A \}. \] 
\item Given $s \in P_{\ofcourse A}^{\infty}$, let $\breve{s}$ be the set of
all $t \in P^{\infty}_A$ such that every finite prefix of $t$ is
$\overline{s(i)}$ for some $ i \in \omega$.  Then, 
\[ W_{\ofcourse A} =  \{ s \in P_{\ofcourse A}^{\infty} \mid \breve{s}
\subseteq W_A \} \]
\end{itemize}
\end{df}

\begin{proposition}

$\ofcourse$ is a comonad on $\Games$, satisfying $\ofcourse = \ofcourse
\circ \ofcourse_w = \ofcourse_w \circ \ofcourse$.  Moreover, $\ofcourse$ has
a natural commutative comonoid structure on its  free algebras, {\em
i.e.} maps 
\[ \delta_A: \ofcourse A \rightarrow \ofcourse A \tensor \ofcourse A \]
such that $\ofcourse$- algebra morphisms between its free algebras are
automatically comonoid homomorphisms.
\end{proposition}
As a consequence of this proposition, the contraction rule is valid in the
game semantics:
\[ \mytwotrans{ \entails \Gamma,  \whynot A, \whynot A} { \entails \Gamma,
\whynot A}  \]
where $\whynot$ is the monad defined by duality from $\ofcourse$: $\whynot
A = (\ofcourse A^{\perp})^{\perp} $.

\subsection{Additives}
The additives of Linear Logic are problematic.  This is seen in various
ways:  by the difficulties of getting a ``reasonable'' implementation  (for
example, in terms of interaction nets) of the commutative conversions for
the additives~\cite{GAL92}; and, most conspicuously, by the problems
they engender with the Geometry of Interaction~\cite{TGI,GI1,GI2}.  

Our notion of polarities throws some light on these matters and suggests a
refinement of Linear Logic which may allow these problems to be addressed.

\begin{proposition}
\posg{\Games} has coproducts, and \negg{\Games} has products, both defined by
disjoint union of games.
\end{proposition}
These definitions can be extended to get weak products and coproducts on
\Games, defined as follows.  
\begin{eqnarray*}
M_{A \with B} &=& M_A + M_B + \{ \ast,l,r \} \\
\lambda_{A \with B} &=& [\lambda_A,\lambda_B,\{ (\ast,P),(l,O),(r,O) \}]
\\
P_{A \with B}&=& \; \mbox{prefix closure of} \; (P_{\negg{A}} + P_{\negg{B}}) \cup  (\ast \cdot l \cdot
P_{\posg{A}} + \ast \cdot r \cdot P_{\posg{B}}) \\
W_{A \with B} &=& (W_{\negg{A}} + W_{\negg{B}}) \cup (\ast \cdot l \cdot
W_{\posg{A}} + \ast \cdot r \cdot W_{\posg{B}})
\end{eqnarray*}
Note that $\negg{(A \with B)} = \negg{A} + \negg{B}$ (disjoint union of
games), {\em i.e.} the weak product in \Games\ is carried to the product in
\negg{\Games} by the co-reflection.

It is important to note that the above proposition is stated only for
\Games, not for $\Games_{\tt hf}$.
History free strategies do {\em not} suffice for the additives.  This seems to
the key reason underlying the problems encountered with additives in the
Geometry of Interaction.

We also note that the surjective pairing axiom for product (and hence the
commutative conversion for With) will only be valid in \negg{\Games}.  This
suggest a syntactic restriction on the With rule, based on the polarities.

Firstly we give a table of how the connectives act on polarities.  Read
$+1$ ($-1$) as ``{\em must be} positive (negative)'' and $0$ as ``{\em may
be} neutral''.
\begin{center}
\begin{tabular}{lll}
\begin{tabular}{||c|c|c|c|c|c|c||}\hline 
$A$ &$B$& $A \tensor B$ & $A \parc B$ & $A \linimpl B$ & $A \with B$ & $A
\aor B$ \\ \hline \hline 
$+1$ & $+1$& $+1$ & $+1$ &  $\;\;\hspace{1mm}0$ & $+1$ & $+1$ \\ \hline 
$+1$ & $\;\;\hspace{1mm}0$& $\;\;\hspace{1mm}0$ & $\;\;\hspace{1mm}0$ & $\;\;\hspace{1mm}0$ & $\;\;\hspace{1mm}0$ & $\;\;\hspace{1mm}0$ \\ \hline 
$+1$ & $-1$ & $\;\;\hspace{1mm}0$ & $\;\;\hspace{1mm}0$ & $-1$ & $\;\;\hspace{1mm}0$ & $\;\;\hspace{1mm}0$ \\ \hline
$\;\;\hspace{1mm}0$ & $+1$& $\;\;\hspace{1mm}0$ & $\;\;\hspace{1mm}0$ & $\;\;\hspace{1mm}0$ & $\;\;\hspace{1mm}0$ & $\;\;\hspace{1mm}0$ \\ \hline
$\;\;\hspace{1mm}0$ & $\;\;\hspace{1mm}0$& $\;\;\hspace{1mm}0$ & $\;\;\hspace{1mm}0$ & $\;\;\hspace{1mm}0$ & $\;\;\hspace{1mm}0$ & $\;\;\hspace{1mm}0$ \\ \hline
$\;\;\hspace{1mm}0$ & $-1$& $\;\;\hspace{1mm}0$ & $\;\;\hspace{1mm}0$ & $\;\;\hspace{1mm}0$ & $\;\;\hspace{1mm}0$ & $\;\;\hspace{1mm}0$ \\ \hline
$-1$ & $+1$& $\;\;\hspace{1mm}0$ & $\;\;\hspace{1mm}0$ & $+1$  & $\;\;\hspace{1mm}0$ & $\;\;\hspace{1mm}0$ \\ \hline
$-1$ & $\;\;\hspace{1mm}0$& $\;\;\hspace{1mm}0$ & $\;\;\hspace{1mm}0$ & $\;\;\hspace{1mm}0$ & $\;\;\hspace{1mm}0$ & $\;\;\hspace{1mm}0$ \\ \hline
$-1$ & $-1$& $-1$ & $-1$ & $\;\;\hspace{1mm}0$  & $-1$ & $-1$ \\ \hline 
\end{tabular}
& \hspace{.8in}
\begin{tabular}{||c|c|c|c||}\hline 
$A$ &$\ofcourse A$& $\whynot A  $ & $A^{\perp}$ \\ \hline \hline 
$+1$ & $-1$& $+1$ & $-1$  \\ \hline 
$\;\;\hspace{1mm}0$ & $-1$& $+1$ & $\;\;\hspace{1mm}0$ \\ \hline
$-1$ & $-1$& $+1$ & $+1$  \\ \hline
\end{tabular}
\end{tabular}
\end{center}
Using these tables as a {\em definition}, we now have a syntactic notion of
polarity, and can use it for the following refined With Rule:

\[ \mytwotrans{ \entails \posg{\Gamma},  A \;\;\;\;\;\entails
\posg{\Gamma}, B} { \entails \posg{\Gamma},  A \with
B} \;\;\;\;\;\;\; \;\;\;\;\;\;\;\;(\mbox{ With}^p )\]

The \posg{\Gamma} is meant to indicate the constraint that all formulas in
$\Gamma$ must be positive.  Let ${\tt LL}^p$ be the modification of
Classical Linear Logic obtained by replacing the usual With Rule with
$\mbox{With}^p$.  Then the commutative conversion for $\mbox{With}$ will be
valid in our game semantics for ${\tt LL}^p$.  We also expect that
${\tt LL}^p$ can be used to extend the Geometry of Interaction
interpretation to the additives.  

\begin{proposition}
There are isomorphisms $\ofcourse(A \with B) \cong \ofcourse A \tensor
\ofcourse B$, $\ofcourse \top \cong {\bf 1}$ and hence ({\em
cf.}~\cite{Seeley}), the co-Kleisli category $K_{\ofcourse}(\Games)$ is cartesian closed.
\end{proposition}

\section{Related Work}\label{relatedwork}
Since a number of researchers have recently examined categories of games,
or at least categories with some game-theoretic flavour, it seems
worthwhile to make some explicit comparisons.

\subsection{Conway games}
As far as we know, the first person to make a category of games and winning
strategies was Joyal~\cite{Joy77}.  His category was based on Conway
games~\cite{Con76} with Conway's addition of games as the tensor product.
Conway's formalization of games differs from ours in that he presents
the tree of positions directly, rather than via an underlying set of moves.
This means that strategies must be formalized as functions on positions,
and hence are necessarily history-sensitive; the possibility of
introducing history-free strategies in our sense does not even arise.

More precisely, a Conway game can be taken to be one of our games with
the following property: for all $a \in M_A$ there is a unique $s \in P_A$
such that $s \cdot a \in P_A$. Call such a game {\em positional}.

\begin{proposition}
Given any game $A$ in $\Games$,  there is a positional game $A^{\rm pos}$
such that $A \cong A^{\rm pos}$ in $\Games$. Moreover, {\em every}
strategy in $A^{\rm pos}$ is history-free. However, $A$ is {\em not}
isomorphic to $A^{\rm pos}$ in $\Games_{\tt hf}$.
\end{proposition}
Thus working with positional games as Conway does would obliterate the
distinction between history-free and history-sensitive which is crucial
to our Full Completeness Theorem. In this respect, our games are more general 
than Conway's.

In another respect, however, Conway games are more general than ours,
at least superficially.  Think of the set
of positions of the game as a tree, with arcs $s \rightarrow s \cdot a$
labelled P or O, according to the label of $a$.  Say that a node is {\em
pure} if all outgoing arcs have the same label, and {\em mixed} otherwise.
In Blass' games, all nodes are pure.  In Conway's games, all nodes are
allowed to be mixed.  Our games are intermediate in generality; the root is
allowed to be mixed, but all other nodes are pure.  Conway games---or their
generalization to the non-positional case---can be
represented in our framework by dropping the stipulation that positions be
{\em strictly alternating} sequences of moves.  His notion of ``sum of
games'', which is used by Joyal as the basis for his construction of a
category of games, then arises by dropping the stipulation from our
definition of tensor product that only Opponent is allowed to switch
components.  This immediately obliterates the distinction between Tensor
and Par; Hyland~\cite{Hyl90} has shown that Joyal's category does not admit
satisfactory interpretations of the additives and exponentials.

Our games are {\em apparently} less general than Conway's; however, as soon
as our definition of tensor product is adopted (with the consequent notion
of morphism; note that Joyal's definition of winning strategy agrees with
ours), this difference disappears.  The key observation is the
following.  Let $A,B$ be Conway games.  Apply our definition of tensor
product to form $A \tensor B$.  Now, because of the stipulation that only
Opponent can switch components, a strictly alternating sequence of moves in
$A \tensor B$ must project onto strictly alternating sequences in $A$ and
$B$.  (Of course, this property fails with Conway's sum of games).  As a
consequence of this, we have the following Proposition.
\begin{proposition}
Let ${\cal C}$ be the category of Conway games, with our definition of
tensor product, and the consequent notion of morphism from $A$ to $B$ as a
winning strategy in $A \linimpl B = (A \tensor B^{\perp})^{\perp}$.  (So,
in particular, this is not the category studied by Joyal~\cite{Joy77}.)
\Games\ is a full subcategory of ${\cal C}$.  If $A$ is a Conway game,
let $A^{\tt alt}$ be the game in \Games\ obtained by deleting all
non-strictly-alternating sequences in $P_A$ (and correspondingly pruning
$W_A$).  Then $A \cong A^{\tt alt}$ in ${\cal C}$; so $\Games \simeq
{\cal C}$.  Moreover, $(A \tensor B)^{\tt alt} \cong A^{\tt alt}
\tensor B^{\tt alt}$.
\end{proposition}
The upshot of this Proposition is that, once our definition of tensor
product---which has been justified both conceptually and by our results in
this paper---is adopted, then one may as well work in \Games\ as in
${\cal C}$.

\subsection{Abstract Games}
De Paiva has studied the Dialectica Categories DC, and Linear categories
GC~\cite{Pai89}.  These are abstract constructions, but reflect some
game-theoretic intuitions. Indeed, Blass applies his game semantics to
DC~\cite{Bla92}.  Again, Lafont and Streicher~\cite{LafStr91} have
developed a ``Game Semantics for Linear Logic''.  An object in the category
${\tt Game}_{K}$ is a structure $(A^{\star}, A_{\star},e)$, where $e: A^{\star}
\times A_{\star} \rightarrow K $, for some fixed set $K$.  If we think of
$A^{\star}$ as strategies for Player, $A_{\star}$ as counter-strategies and
$e$ as the payoff function, we see some connection with game-theoretic
ideas.  However, this model is very abstract; in fact it forms a
particular case of Chu's very general construction of $\star$-autonomous
categories from symmetric monoidal closed categories~\cite{Barr79}.

In summary, these models have only rudimentary game-theoretic content and hence
only a very weak relation with our work.

\subsection{Blass' game semantics}
Blass' game semantics for Linear Logic is by far the nearest
precursor of the present work.  While we happily acknowledge its
inspiration, we must also say that, in our opinion, our semantics is a
decisive improvement over that of Blass, as our results show.

It is worth setting out the key points in some detail, since our
identification of the problems in Blass' semantics was a crucial step in
our own work and differs sharply from Blass' analysis of the
discrepancy between his semantics and Linear Logic.

The games Blass considers correspond to those in $\posg{\Games} \cup
\negg{\Games}$ in our framework; that is, to either positive games (all
opening moves by Player) or negative games (all opening moves by Opponent).
This means, among other things, that all connectives must be defined by
cases on the polarity of their arguments; and, more importantly, the
resulting game must itself have a definite positive or negative polarity.
The plays in Blass' games are then started by Player for a positive game
and by Opponent for a negative game.

The key difference between Blass' approach and ours concerns the definition
of tensor product.  Blass' rule for who moves next in the tensor product is
that Player moves if he is to move in either game.  This makes sense if we
think of ``Opponent to move'' as a kind of approximation to the proposition
represented by the tensor product being true---since the onus is on the
Opponent to move in order to avoid defeat---and the tensor as a kind of
conjunction.  Surprisingly enough, this definition turns out to 
{\em almost} coincide with ours.
Suppose that we are in a position where Opponent is to move in both subgames;
then he has the choice of moving in either component, leading to a position
where Player is to move in just one component. In this latter situation,
Player is forced to move in the component where Opponent last moved.
Such a move will return us to a situation where Opponent is to move in both
components.
This leaves just one anomalous situation, where Player is to start in both
components.
This is the only case
where the situation can arise that Player must move next in both games.  Note
that in our framework, this situation can never arise at all.  Also, note
that this situation contradicts our previous analysis of tensor; for
example, in terms of the trip conditions, it corresponds to the forbidden
sequence $\ans{A}\ans{B}$.  Blass treats this anomalous situation as a
special case; Player makes his opening move simultaneously in both
components.  This special case is at the heart of the pathologies in his
semantics.

\subsubsection{Composition}

Composition is not associative in Blass' semantics~\cite{Blap92}; so
he does not get a category of games at all.  

{\bf what to say here}

Define games $A,B,C,D$ as follows:
\begin{eqnarray*}
A &=& (\{ a \}, \{ (a, O) \},\{   \epsilon, a  \},\varnothing) \\ 
B &=& (\{ b_1,b_2 \}, \{ (b_1, P),
(b_2,O) \},\{  \epsilon, b_1, b_1 \cdot b_2\}, \varnothing) \\ 
C &=& (\{ c \}, \{ (c, O) \},\{   \epsilon , c\}, \varnothing ) \\
D &=& (\{ d \}, \{ (d, P) \},\{  \epsilon, d \}, \varnothing)
\end{eqnarray*}
Here a move $\langle b_1, b_2 \rangle $ is an opening move in the special case
described above.

There are winning strategies $\sigma: A \rightarrow B, \; \tau: B
\rightarrow C, \; \upsilon: C \rightarrow D$.  $\sigma$ is the strategy
that forces the entire play to stay in constituent $A^{\perp}$ after the
first move.  Similarly, $\upsilon$ is the strategy that forces the entire
play to stay in constituent $D$.  $\tau$ is the strategy that responds to
the initial move of the Opponent with a move in $B$.  More precisely,
\begin{eqnarray*}
\sigma &=& \{ \epsilon, a \}  \\
\tau &=& \{ \epsilon, \langle b_1,c \rangle , \langle b_1,c \rangle \cdot
b_2  \} \\ 
\upsilon &=& \{ \epsilon , d \}
\end{eqnarray*}
Thus,
\begin{eqnarray*}
(\sigma ; \tau) ; \upsilon &=& \upsilon \\
\sigma ; (\tau ; \upsilon) &=& \sigma 
\end{eqnarray*}
and hence unequal.

\subsubsection{Weakening}
Weakening is valid in the Blass semantics.  To see why, suppose that
Player has a winning strategy for $\Gamma$.   Consider the game $\Gamma, A$.
If $A$ is positive, Opponent cannot move in $A$ and since only Player
can switch components in a Par, we need never play in $A$ at all.  (Of
course, this is exactly the argument for the validity of weakening with
respect to $\whynot_w A$ in our semantics).  If $A$ is negative, there are
two cases.  

\begin{itemize}
\item Some game in $\Gamma$ is positive: so Player is to start
in $\Gamma$ and $\Gamma, A$.  Thus, Player can simply play his strategy for
$\Gamma$ without ever entering $A$.  

\item All games in $\Gamma,A$ are negative: The special case takes effect
and Opponent must make his opening move in every component of $\Gamma,A$.
Then, Player can simply ignore the opening move in $A$ and play as he
would have done in response to the opening moves in $\Gamma$.

\end{itemize}
By contrast, in our interpretation, unless $A$ is positive, Opponent can
move in $A$, and Player may have no way to respond; so Weakening is not
valid.

\subsubsection{An Example}
Consider the example discussed in Blass' paper (\cite{Bla92}, pp.210-213).
The sequent considered there is:

\[ (A^{\perp} \parc B^{\perp}) \tensor (C^{\perp} \parc D^{\perp}), (A \parc
C) \tensor (B \parc D) \]
We describe a  strategy for Opponent, which with suitable choice of games
for $A, \;B,\; C, \;D$ will defeat Player in our semantics.
\begin{enumerate}
\item Opponent moves in $A$.
\item Player moves in $A^{\perp}$.
\item Opponent moves in $C^{\perp}$.
\item Player moves in $C$.
\item Opponent moves in $B$.
\end{enumerate}
At this point, Player needs to move in $B^{\perp}$; however, he cannot,
because it is Opponent's move in the sub-game $A^{\perp} \parc B^{\perp}$.
What saves the Player in Blass' semantics is again the special case, which
would force Opponent to move in both $B$ and $D$ simultaneously, thus
allowing Player to respond in $D^{\perp}$.

\subsection{Sequential Algorithms}
Lamarche~\cite{Lam92} and more recently, but independently,
Curien\footnote{Curien's email announcement of his results appeared
following ours~\cite{Abr92b} announcing the results of this paper.}~\cite{Cur92} have found linear decompositions of the Berry-Curien
category of sequential algorithms on (filiform) concrete data
structures~\cite{BC85}.  That is, they have described models of Linear
Logic (Intuitionistic Linear Logic only, in Curien's case) such that the
co-Kleisli category is equivalent to the Berry-Curien category.  Moreover,
these Linear categories have a game-theoretic flavour.  In fact, we have
the correspondence:

\begin{center}
\begin{tabular}{||l|l||}\hline 
Game & Concrete Data Structure  \\ \hline
O-Moves & Cells  \\ \hline
P-Moves & Values \\ \hline
Positions & Enabling Relation \\ \hline
Strategy  & State\\ \hline
\end{tabular}
\end{center}

We have not seen the full details of Lamarche's work; Curien's construction
can be related to our work as follows.  The objects in his category are
exactly our negative games, minus the information about infinite plays.
The morphisms correspond to strategies---which need be neither
history-free nor winning. His interpretations of the Intuitionistic linear
connectives, with these provisos, appear to correspond to ours.  We take
this link with sequential algorithms as an encouraging confirmation of the
potential of game semantics.  We note, finally, that the connection between
sequential algorithms and negative games confirms our identification of
negative games with demand-driven computation.  This also ties up with the
first author's association of $\with$ and $\ofcourse$ (more precisely of
$\ofcourse_w$) with lazy evaluation~\cite{AbrCIL91}.

\addcontentsline{toc}{section}{\protect\numberline{}{References}}
\bibliography{concurrency,dfbib,biblio}
\end{document}